\documentclass[10pt,journal,compsoc]{IEEEtran}

\usepackage{cite}
\usepackage{amsmath,amssymb,amsfonts}
\usepackage{algorithm}
\usepackage[noend]{algorithmic}
\usepackage{graphicx}
\usepackage{textcomp}
\usepackage{xcolor}
\usepackage{url}
\usepackage{hyperref}
\usepackage{subcaption}
\usepackage{soul}
\usepackage{enumitem}
\definecolor{BLUE}{rgb}{0,0,1}

\newcommand{\comment}[1]{}

\begin{document}

\title{CoANE: Modeling Context Co-occurrence for Attributed Network Embedding}

\author{
        I-Chung~Hsieh,
        Cheng-Te Li,~\IEEEmembership{Member,~IEEE}
\IEEEcompsocitemizethanks{\IEEEcompsocthanksitem I-Chung Hsieh, Institute of Data Science, National Cheng Kung University, Tainan, Taiwan.
\IEEEcompsocthanksitem Cheng-Te Li, Institute of Data Science, National Cheng Kung University, Tainan, Taiwan.
}
\thanks{
}}

\markboth{Journal of \LaTeX\ Class Files,~Vol.~14, No.~8, August~2015.}%
{Shell \MakeLowercase{\textit{et al.}}: Bare Demo of IEEEtran.cls for Computer Society Journals}

\IEEEtitleabstractindextext{%
\begin{abstract}
Attributed network embedding (ANE) is to learn low-dimensional vectors so that not only the network structure but also node attributes can be preserved in the embedding space. Existing ANE models do not consider the specific combination between graph structure and attributes. While each node has its structural characteristics, such as highly-interconnected neighbors along with their certain patterns of attribute distribution, each node's neighborhood should be not only depicted by multi-hop nodes, but consider certain clusters or social circles. To model such information, in this paper, we propose a novel ANE model, \textit{Context Co-occurrence-aware Attributed Network Embedding} (CoANE). The basic idea of CoANE is to model the context attributes that each node's involved diverse patterns, and apply the convolutional mechanism to encode positional information by treating each attribute as a channel. The learning of context co-occurrence can capture the latent social circles of each node. To better encode structural and semantic knowledge of nodes, we devise a three-way objective function, consisting of positive graph likelihood, contextual negative sampling, and attribute reconstruction. We conduct experiments on five real datasets in the tasks of link prediction, node label classification, and node clustering. The results exhibit that CoANE can significantly outperform state-of-the-art ANE models.
\end{abstract}

}

\maketitle

\IEEEdisplaynontitleabstractindextext

\IEEEpeerreviewmaketitle

\IEEEraisesectionheading{\section{Introduction}\label{sec:introduction}}

\IEEEPARstart{N}{etworks} are important data structures to represent the relationships between entities. Modern techniques in Web, storage, and computation allow us to process, retrieve, and discover knowledge from a variety of network data. In the real world, for example, social networks depict the relationships and interactions between people, and academic citation networks encode how papers are referred to each other. In addition to the graph structure, there are usually attributes associated with nodes in the networks. In social networks, such as Facebook and Instagram, users can maintain their profiles. In academic citation networks, such as Google Scholar, researchers possess affiliation, expertise, and profiles. Jointly modeling the network structure and node attributes can benefit applications, such as recommender systems~\cite{Grbovic-Cheng:KDD-2018} and fake news detection~\cite{Liu-Wu:AAAI-2018}. 

Network embedding (NE) is an essential technique in various network mining and prediction tasks ~\cite{nesurvey18}. The basic idea is to learn low dimensional feature representation vectors of nodes so that the graph neighborhood of each node can be encoded in the feature space. The performance of typical tasks, such as link prediction, node label classification, and community detection, can get improved based on NE. Several typical NE models were proposed, such as DeepWalk~\cite{Perozzi-et-al:KDD-2014}, LINE~\cite{Tang-et-al:TANG-2015}, and node2vec~\cite{Grover-Leskovec:KDD-2016}. The general idea is to generate contexts of nodes and apply the skip-gram model~\cite{mikolov2013efficient} to learn and produce the embedding vectors. However, typical models focus on utilizing network structure, but attributes associated with nodes are neglected. 

The recent focus shifts to attributed network embedding (ANE), whose goal is to preserve not only the network structure but also node attributes when learning embeddings. 
GAT2VEC~\cite{sheikh2018gat2vec} transforms node-attribute relations into a bipartite graph and merges the embeddings of structure and attributes based on random walk and skip-gram model. LANE~\cite{huang2017label} incorporates all features by adding node labels and preserves their correlation. NEEC~\cite{huang2018exploring} concentrates on attributed network learning with expert cognition that requires queries and answers from the experts to improve embedding. The non-linear mapping with random walk process is also developed for ANE in UPP-SNE~\cite{zhang2017user}. SANE~\cite{wang2018united} further imposes an attention mechanism to discriminate the correlation between nodes. The state-of-the-art methods are DANE~\cite{Gao-Huang:IJCAI-2018} and ASNE~\cite{Liao-et-al:IEEE-2018}. DANE captures the high non-linearity and preserves various proximities in both topological structure and node attributes. ASNE preserves both structural proximity and attribute proximity so that the global structure and the homophily effect in attributes can be captured.

Although several attributed network embedding models were proposed, we find that they all focus on either learning representation of network structure or attributes individually, or find a proper approach to combine the feature representations of such two parts. In this way, the target node's diverse aspects between network structure and attributes cannot be modeled. For example, in real-world networks like a social network, a node links to a number of nodes, and has multi-hop neighboring nodes consisting of several egocentric communities (i.e., the so-called \textit{social circles}), such as family and school, in which nodes are tightly interconnected and surely have attributes similar to each other. And those belonging to one or more distinct communities can be differed from not only topological connections in their neighborhood, but also their attributes. Only when we jointly model network structure and attributes together, we will be able to better exploit their underlying correlation. State-of-the-art methods GAE~\cite{Kipf-Welling:arXiv-2016} and VGAE~\cite{Kipf-Welling:arXiv-2016} can simultaneously model network structure and attributes. They scan the first- or second-order graph neighborhoods recursively so that the further and wider order region in the network can be indirectly reached and exploited. Then all neighbors in the same order would be considered to have the same importance for the target node. 
We think that without better use of connections between neighbors in a fine-grained manner, it is less possible to distinguish same-hop neighbors from each other, which could belong to different \textit{latent social circles}. For instance, neighbors of a student may have multiple social circles, such as ``CS dept'', ``family'', and ``labmates'', whose sizes are different from one another. Friends from ``CS dept'' circle with dominated attributes would dilute the information on ``family'' circle. Besides, an effective embedding learning model needs to better exploit wider and deeper interconnected neighbors of multiple latent social circles.

This paper aims to capture specific contexts representing the latent social circles of the target node and to leverage \textit{context co-occurrence in both network structure and node attributes} for better embedding learning even though some edges are missing or unobserved. 
In other words, the co-occurrence of node contexts in both network structure and node attributes provides information about which neighboring nodes and their relative attributes are correlated in the context. 
Comparing to existing studies~\cite{Gao-Huang:IJCAI-2018,Liao-et-al:IEEE-2018,zhang2018anrl},
ASNE~\cite{Liao-et-al:IEEE-2018} simply treats attributes as the model input to predict the representation of a node. 
Though DANE~\cite{Gao-Huang:IJCAI-2018} additionally considers fusing preservation, and it results in higher optimization costs and more complex architectures. 
ANRL~\cite{zhang2018anrl} leverages common neighbors derived from random walk for topology preservation, but it does not model node attributes with the context co-occurrence structure.
Therefore, we think that the better modeling of the interplay between network structure and node attributes can enhance the embedding power.

To deal with these issues, in this paper, we develop a novel ANE model, \textit{Context Co-occurrence-aware Attributed Network Embedding} (CoANE) \footnote{The code of CoANE can be accessed via the following Github link:
\url{https://github.com/ICHproject/CoANE/}
}, to learn node embeddings in attributed networks. Our CoANE is devised to make the embeddings of nodes preserve three-fold information: (a) the graph neighborhood of nodes from sampling contexts: making nodes tightly interconnected have similar embeddings, (b) the context co-occurrences of nodes: driving the embeddings to become closer if nodes whose direct and indirect neighbors (i.e., contexts) share similar attributes, and (c) the attributes of nodes: shaping the embeddings of nodes to preserve their own attribute information.
To realize such ideas, CoANE is developed to have three novel components.
First, we generate the crucially structural contexts of nodes via random walk and treat them from the perspective of attributes. 
Second, by considering each attribute as a \textit{channel}, we adopt a 1-D convolutional layer with different filters to model contexts with their corresponding channels and summarize the features from the target's contexts into the derived embedding.
Third, for more effective learning, we extend two contextual likelihood approaches in the design of our objective function. 
One is the \textit{positive graph likelihood} whose goal is to make one- and high-order structural context co-occurrences be better preserved in the embeddings. 
The other is a \textit{contextually negative sampling}, which considers the co-occurrence frequency of the target node and the negative samples to have a more effective negative sampling. 
Besides, we also perform the reconstruction of attribute values based on the derived node embeddings so that the semantics of nodes can be better preserved.

%

%
We summarize the contributions of this paper as follows.
\begin{itemize}[leftmargin=*]
  \item We propose a novel ANE model, CoANE, which is able to better generate and model the representational contexts of the target node. The main idea of CoANE is to capture the latent social circles of the target node. 
  
  \item Technically, a convolutional mechanism is applied to distill positional information and latent social-circle features from the attributed contexts, which cannot be captured by existing solutions that use fixed-hop neighborhood. 
Besides, the extended graph likelihood, the contextually negative sampling method, and the reconstruction of attributes are also proposed for preserving higher-order structural relationships and nodes' semantic knowledge.

  \item Experiments conducted on five real datasets in tasks node classification, node clustering, and link prediction show that CoANE can significantly and consistently outperform state-of-the-art ANE methods.
\end{itemize}

This paper is organized as below.
We discuss relevant studies in Section 2, and present the technical details of CoANE in Section 3. The experimental settings and results are described in Section 4. Section 5 concludes this work.
%
\section{Related Work}
We first introduce random walk-based methods that are popularly adopted for network embedding to attributed network embedding (ANE) in Section 2.1. Then, we discuss well-known deep graph reconstruction approaches to ANE in Section 2.2. We point out that in the literatures of network embedding, the techniques of subgraph aggregation can better incorporate network structure with node attributes in Section 2.3. Last, in Section 2.4, we review the common optimization designs for the ANE methods discussed from Section 2.1 to 2.3, and point out their insufficiencies.

\subsection{Random Walk-based Approaches}
To review the random walk-based methods for network embedding,
we first introduce DeepWalk~\cite{Perozzi-et-al:KDD-2014}, which generates fixed-length paths to have neighboring correlated nodes, and then computes the embedding similarity between the center node and a random-selected node inside/outside the window. Then, node2vec~\cite{Grover-Leskovec:KDD-2016} devises the biased random walk to balance the wide and deep neighbors. 
Both approaches can transform the network structure to shortly and tightly interconnected sentence-like paths. 
Recent methods incorporate deep learning with random-walk node sequences.
NetRA~\cite{yu2018learning} presents a generative adversarial training and produces node embeddings by reconstructing random walk sequences and discriminating positive and negative samples. 
However, DeepWalk, node2vec, and NetRA are not designed for ANE. 
Thus, STNE~\cite{liu2018content} utilizes seq2seq in machine translation. It considers the random walk sequence as a sentence and learns the higher-order information. 
GraphRNA~\cite{huang2019graph} conducts a random walk on the bipartite network between nodes and attributes, and then learns the features from the mixed context via sequence modeling for node classification. 
Besides, ANRL~\cite{zhang2018anrl} combines the skip-gram model as an additional reconstruction for node attributes' autoencoder model with a jointly-learning process. 
Although obtaining the promising performance, both STNE and GraphRNA ignore the distribution of sequence sampling and node context diversity. ANRL only applies the distribution to the information preservation, and similar attributes in the context are not discussed.
For this issue, we find that metapath2vec~\cite{m2v} can learn the node embeddings considering various relational patterns of nodes and edges, but it requires a heterogeneous network as the input graph. 
The strength of metapath2vec is on modeling diverse types of relationships between nodes, rather than incorporating node attributes.
Hence, the fine-grained and diverse semantics distributed over contexts cannot be encoded into node embeddings.
\vspace{-5pt}

\subsection{Graph Reconstruction-based Approaches}
For deep learning applications, the autoencoder-based methods get attention, in which the input is squeezed into the embeddings and maintains the important features by optimizing the error between the original input and the decoder's output. 
To extend to the ANE framework, ASNE~\cite{Liao-et-al:IEEE-2018} learns the embeddings from different information sources and feeds them into the multi-layer neural network to distill high-level features.
Then, DANE~\cite{Gao-Huang:IJCAI-2018} adopts both the multi-layer mapping and the complementary loss to enforce two embeddings being as consistent as possible.
SCAN~\cite{meng2019semi} further considers combining diverse knowledge, including graph structure, node features, and node labels, for ANE in a semi-supervised setting. 
Although autoencoder-based methods can produce promising better performance, neither diverse contexts nor context importance is encoded in the embedding learning. 
Besides, there is a trade-off between performance and computation cost, especially for sparse features (e.g., the adjacency matrix), and the multi-source information fusion.
Recent studies argue the incompatibility of the network embedding on Euclidean space for the structure of real-world networks.
%
To this end, DRNE~\cite{tu2018deep} further incorporates the regular equivalence into node embedding learning, which is proven to preserve some typical node centrality measures.
%
\vspace{-5pt}

\subsection{Subgraph Aggregation-based Approaches}
To better incorporate structure and attribute, the technique of subgraph aggregation is used in graph representation learning. The main idea is to utilize the neighboring subgraph to represent each node, and to learn an aggregation function that can fuse features of neighboring nodes with the target node. To fulfill the subgraph aggregation technique, since the convolutional mechanism has been a powerful detector of local features, especially for relationship pattern recognition. 
Graph Convolutional Network (GCN)~\cite{kipf2016semi} is proposed to adopt the spectral graph convolution for semi-supervised node classification. Its extended method Graph Attention Network (GAT)~\cite{velivckovic2017graph} can further distinguish the importance of neighbors. 
Nevertheless, we concentrate on node embedding learning in attributed networks without known labels. 
For the unsupervised setting, Graph Auto-Encoder (GAE)~\cite{Kipf-Welling:arXiv-2016} and Variational Graph Auto-Encoder (VGAE)~\cite{Kipf-Welling:arXiv-2016} revise the GCN and learn ANE by adding the reconstructed objective of the adjacency matrix, along with variational parameters for the probability function. 
Furthermore, GraphSAGE~\cite{hamilton2017inductive} is devised for learning both \textit{transductive} and \textit{inductive} node embeddings in a graph.
In GraphSAGE, node features, along with the graph topology, are used to learn an embedding function that can be applied to both existing and new-coming nodes in the graph. GraphSage can be trained in either semi-supervised or unsupervised manners.
To enhance node embeddings' quality and robustness, ARGA/ARVGA~\cite{pan2018adversarially} extends the GAE/VAGE by the adversarial learning to generate more real negative samples for model learning.

However, these unsupervised subgraph aggregation models cannot distinguish the contributions of different hops of neighbors, and cannot capture the structural and attributed patterns, such as social circles. 
Moreover, their objective functions make embeddings fit the input adjacency matrix that only considers first-order neighbors. The preservation of higher-order
information tends to be missing.
\vspace{-5pt}

\subsection{Optimization Design} 
In optimizing node embedding learning, the general design of loss function consists of two parts: (1) enhancing the similarity of nodes that are close and correlated with each other, and (2) using negative sampling to better separate irrelevant nodes from one another. For the first part, encoding the first-order relationship between nodes, such as GAE~\cite{Kipf-Welling:arXiv-2016}, VGAE~\cite{Kipf-Welling:arXiv-2016} and STNE~\cite{liu2018content}, is the common-used way. However, it would result in being less capable of learning the distributions of importance among different neighbors. Preserving higher-order relationships between nodes by DANE~\cite{Gao-Huang:IJCAI-2018} can lead to better performance. 
Second, the negative sampling is to replace the high-cost computation of softmax mapping for large-scale networks. 
However, producing negative samples based on their appearance frequency cannot well separate irrelevant nodes from each other in the embedding space because nodes tend to have similar negative samples.
In this work, we pay more attention to the preservation of topological distribution of the network by incorporating the weighted sampling probability of context into negative sampling.


\begin{figure*}[!t]
  \centering
  \includegraphics[width=0.9\textwidth]{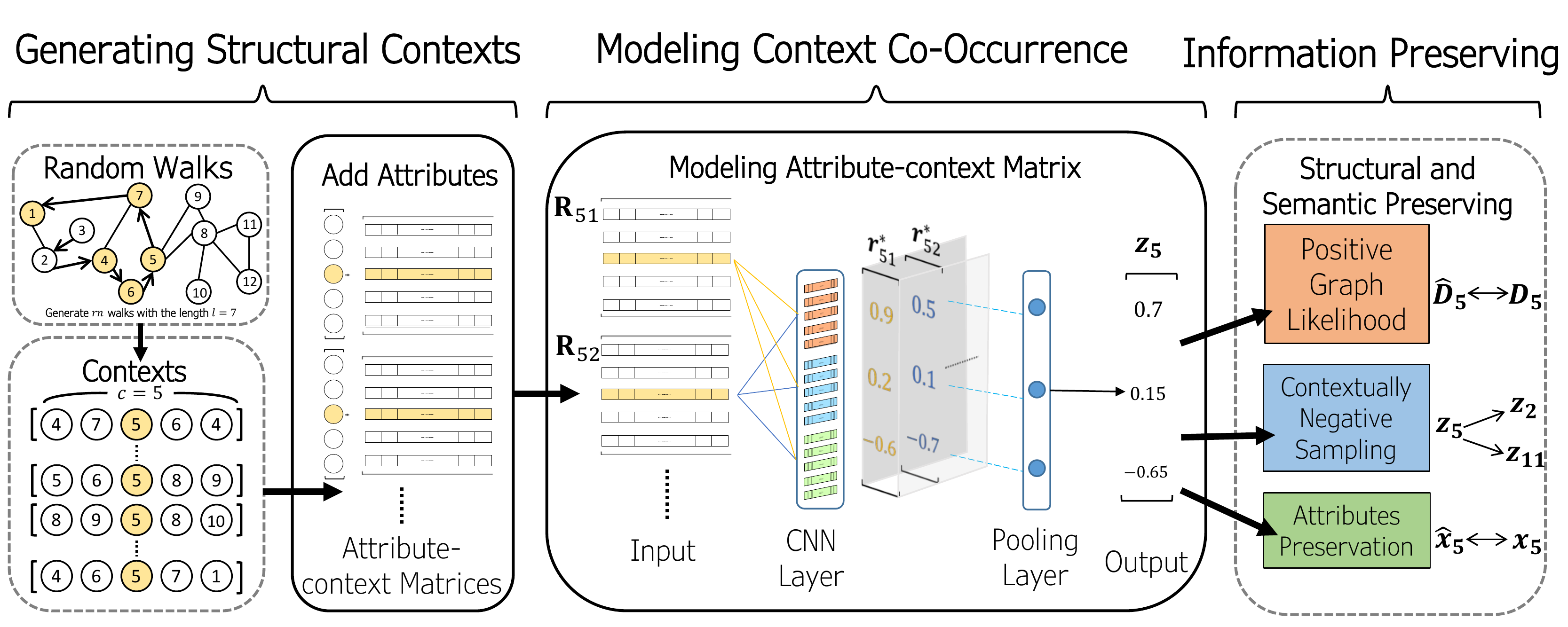}
  \caption{The overview of CoANE. In this illustration, we generate $r$ random walks with length $l = 7$, as indicated by bold directed edges, for $n = 12$ nodes. The contexts of the midst node $5$ are extracted from random walks with size $c = 5$. The attribute-context matrix $\mathbf{R}_{5*}$ is fed into $1$-D convolutional layer and then $1$-D pooling layer. The low-dimensional embedding is the output. For structural and semantic preserving, we update the model parameters by an extended graph likelihood, the proposed contextually negative sampling, and the attribute preservation.}
  \vspace{-10pt}
  \label{fig:architecture}
\end{figure*}

\vspace{-5pt}

\section{The Proposed CoANE Model}
We first describe the notations. Given an attributed network $\mathbf{G} =(V,\mathbf{E},\mathbf{X})$, where $V$ is the node set with $n$ nodes ($|V| = n$), $\mathbf{E}$ is the adjacency matrix, in which $\mathbf{E}_{ij}$ represents the weight between $v_i \in V$ and $v_j \in V$, and $\mathbf{X} \in {\rm I\!R}^{n \times d}$ is the node-attribute matrix, where $d$ is the number of attributes.
In the task of NE learning, we aim at generating a low dimensional vector to represent each node in a network. As for ANE, the goal is to preserve two properties from an attributed network, \textit{structural proximity} and \textit{semantic proximity}. The former indicates that nodes with similar structural neighborhood tend to have similar embedding vectors. The latter aims at making nodes sharing similar attributes possess similar embedding vectors. We think such structural and semantic proximities are correlated with one another in the form of \textit{social circles}, and should be simultaneously modeled in learning node embeddings.
Here the \textit{social circles}~\cite{egonet12} refers to that there are multiple groups of friends that share similar attributes and have tight connections to each other in a node's ego network. The preservation of structural and semantic proximity in CoANE is to capture the latent social circles surrounded by each node in the network. 
The framework of CoANE contains three parts including \textit{generating structural contexts}, \textit{modeling context co-occurrence}, and \textit{proximity preserving}, as shown in Fig. \ref{fig:architecture}. We first find the contexts via random walk and combine them with attributes for modeling their context co-occurrence features. In the end, we compute the likelihood of preserving co-occurrence proximities in updating our model and deriving the resulting embeddings. We elaborate details of our CoANE in the following subsections. 
\subsection{Generating Structural Contexts}
To capture the latent social circles, we start from generating sequences of correlated nodes, i.e., structural contexts, via random walk based on word2vec~\cite{Mikolov-et-al:NIPS-2013}. 
At each walking step, the random walker starts at a given node $v_i \in V$ and decides which adjacent node to visit next by sampling with probability $p(v_i) = \frac{\mathbf{E}_{ij}}{\sum_j \mathbf{E}_{ij}}$. For each node as the starting one, we repeat the sampling process until the length of walks is up to a pre-defined value $l$. We can repeat this process $r$ times for each node, and then have $rn$ sequences with length $l$. 
Then, we compile each node's context by scanning the sequences and copying a specific fragment as a context. We set a fixed window (context) size $c$ and align the midst of the window with the starting node of sequences. 
A context contains the target node's previous and latter neighbors. Since the beginning position of the scanning contains empty slots, i.e., the first half of the window has no previous neighbors, we perform padding to fill in the empty slots like the image padding for the convolutional neural network (CNN). 
Then, we move the window towards the next positions in the sequence, and adopt subsampling~\cite{Mikolov-et-al:NIPS-2013} to alleviate over-emphasizing the nodes that high-frequently appear in the context. 

The subsampling is to deal with the imbalance of occurrence frequency between rare and frequent nodes. To alleviate the over-frequent occurrence for some nodes, those nodes with higher frequency should be ruled out with higher probabilities. That said, in subsampling, we increase the possibility of being sampled for rare nodes, i.e., those rarely appear in the generated node sequences so that the quality of embeddings of rare nodes can be improved.
The subsampling probability is given by $p_{sub}(v) = 1-\sqrt{t/f(v)}$, where $t$ is a constant, and $f(v)$ is the frequency of node $v$'s appearance in the generated node sequences. The context of the midst with the frequency $f(v)$ higher than $t$ would tend to be discarded. In addition, we set $p_{sub}(v) = 1$ if $v$ is the starting node for each sequence to ensure that each node has at least one context neighbor. 

The original skip-gram model~\cite{Mikolov-et-al:NIPS-2013} considers only the center node and its neighbors in the sampled node sequence for embedding learning. Since nodes outside the context window are discarded, the original skip-gram needs to sample more node sequences, which produces additional computational cost. 
We think the entire context (i.e., the sampled node sequence) can reflect the latent social circles that the center node involves. Besides, the distribution of positional information located from near-by to far-away neighbors with respect to the center node can also be used to unfold the sizes of different latent social circles. Hence, we fully exploit all of the generated contexts, and accordingly model the context co-occurrence to learn latent social circles of every node. 

An illustration architecture of our random walk mechanism is shown in the left dotted blocks of Fig. \ref{fig:architecture}. In the random walks block and contexts block, we demonstrate the generated contexts for node $5$, which are sampled from the random walk node sequences. We can find the nodes adjacent to the node $5$ must be the 1-hop, 2-hop, and higher-order neighbors.
In the following, we denote the collection of contexts with the same central node $v$ as a set $context(v)$. Because the generated context comes from the sampling process,
the size of each node $v$'s context set $context(v)$ tends to be different. This indicates that a node with more diverse neighbors is surrounded by richer social circles. Such a setting depicts that a neighbor that frequently occurs in the context can possess similar traits (sharing common neighbors or similar attributes) as the midst node. Hence, we need to learn the specific patterns from the context to distinguish features between nodes.

Note that although the undirected networks do not have positional information, the distance between nodes in the network can indicate the relationship strength between nodes. The higher-order relationship (i.e., long-distance) can help find the latent features. It could also introduce noise if the random walk goes in the wrong way. Therefore, we adopt the convolutional mechanism, in which the filters with context-like length are used, to learn which parts in the high-order relationships contribute more in depicting the context of nodes. The positional information is captured by weights learned from convolution filters that are applied to generated contexts. Different nodes can have various distributions of useful positional information. The positional information can help better represent nodes because high-order relationships in nodes' context can be distinguished by convolutional weights.

Modeling context co-occurrence can capture positional information. We create the \textit{co-occurrence matrix} $\mathbf{D}$ to represent the co-occurrences of nodes by counting the node occurring the contexts of the node $v_i$ (i.e., $\mathbf{D}_{ij} =$ the counts of $v_j$ in all $context(v_i)$ ).
Since one-hop context neighbors of a node can best represent that node, to enhance the preservation of local structure, we also define $1$-hop co-occurrence $\mathbf{D^1}$, where $\mathbf{D^1}_{ij} = \mathbf{D}_{ij}$ if $\mathbf{E}_{ij} >0$.
These two co-occurrence matrices help us preserve the graph information in the optimization step.

\vspace{-4pt}
\subsection{Modeling Context Co-occurrence}
We learn the pattern from each node's context and combine the context features. 
We adopt the convolutional mechanism, along with multiple filters, to extract the similar attitudes and positional information of nodes in a context, and then pool all context features to generate node embeddings. 
Similar to the CNN model for the feature extraction of the image: a cat image can be assembled by several small matrices depicting a long tail, a furry body, and round pupils. 
We can imagine such the context set of the target node is a \textit{picture} consisting of specific features from their nodes' attributes and topological neighbors. 
Since the diversity of contexts for the target needs to be encoded, we need a multi-view model to recognize and extract a variety of patterns of ``pixels'' in different contexts. 


The idea of modeling context co-occurrence is to capture the \textit{latent social circles} surrounded by each node in the network. The social circles mean that in a social network, the neighborhood (e.g., friends and friends of friends) of a user tends to contain multiple neighboring subsets, in which nodes in each subset are tightly connected with one another and share similar attributes. Each neighboring subset is considered as a social circle. For example, in reality, each social circle (i.e., neighboring subset) can be ``basketball club'', ``family'', and ``labmates.'' The convolutional mechanism that learns the combination between graph structure and context attributes can model the latent social circles through a variety of convolutional filters. 
Since the attributes of nodes further away from the midst in the context are less correlated with the target node, we learn different weights by wider convolutional filters to adjust the contribution of different positional information in the context. 

\begin{figure}[!t]
  \centering
  \includegraphics[width=0.95\linewidth]{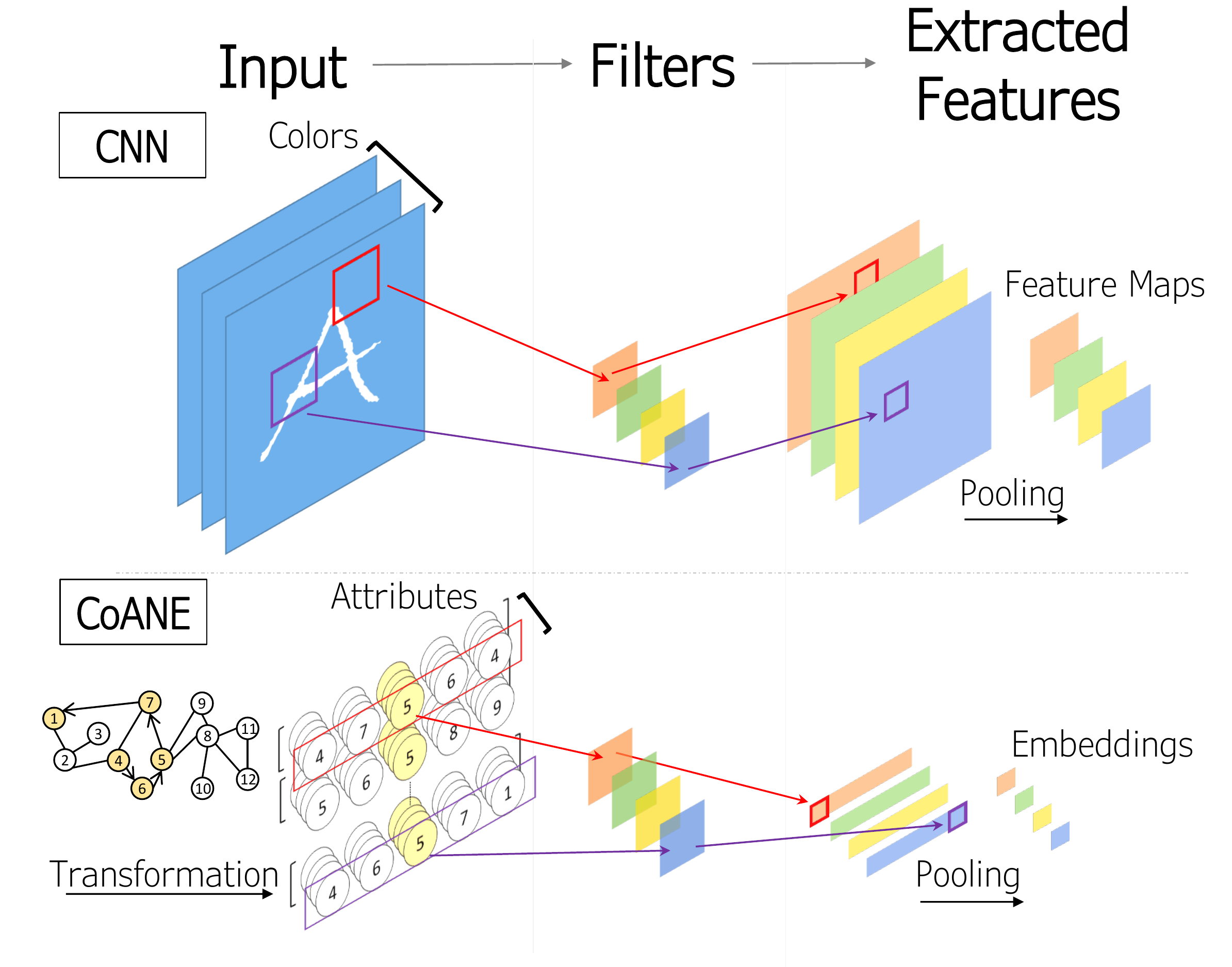}
  \caption{The differences of convolution between CNN and CoANE. After converting the attributed network into a set of attribute-context matrices, each matrix can be viewed as a segment of photo with certain features that we perform convolution through multiple filters.}
  \vspace{-12pt}
  \label{fig:CNNvsCoANE}
\end{figure}

The mathematical details of modeling these contexts are depicted as follows. First, we represent each context for the same central node in the form of the matrix. For every node in the context, we put and align their attributes of these context nodes together according to their order in the context. An \textit{attribute-context matrix} $\mathbf{R}_{vi} \in {\rm I\!R^{c \times d}}$ for the $i$-th context with midst node $v$ can be derived. 
We use all the attribute-context matrices involved by the same midst $v$ to distill its features, where each matrix can be viewed as the source of a feature. 
Specifically, the attribute-context matrices corresponding to the same midst $v$ can be concatenated vertically as a large attribute-context matrix $\mathbf{R}_{v} \in {\rm I\!R^{c' \times d}}$, where $c' = c\cdot|context(v)|$. Its attributes in the second dimension of the matrix are independent. Hence, we can view each attribute as a channel, like the RGB color values of an image, to depict the specific pattern between network structure and attributes. Hence, the matrix can be squeezed along the second dimension as a sequence with $d$ channels with its features. The summarized matrix from these contexts has become a fixed form, which is exactly compatible with the input form of the convolutional neural network in Euclidean space.
Since a sequence of attribute-context matrices is corresponding to contexts per midst node, then we can directly adapt the \textit{$1$-dimension convolutional neural network} (1-D CNN) to learn the similar attributes of nodes at different positions for each context. 
Note that we do not consider the overlapping of the receptive region because each context individually represents the target node.
We let the filter of CNN scan the region in the length of context instead of the overlapping scan. In other words, the setting of our $1$-D CNN model includes: the number of input channels (attribute dimension) $d$, the number of output channel (embedding dimension) $d'$, and the volume (receptive field size) $=$ $ c$ (context size) with stride $= c$ so that each movement of the filter is equal to the length of context as an unit. That is, the model filters the attribute-context matrix $\{x_{v_{-c'}},...,x_{v},...,x_{v_{c'}}\}$ for the context $\{v_{-c'},...,v,...,v_{c'}\}$, where $x_v$ is the corresponding attribute vector of node $v$ and $c' = (c-1)/2$, and they are convolutionally summed by $d'$ filters to be a feature vector $\mathbf{z} \in {\rm I\!R}^{d'}$. After having scanned all of the sequences, we can derive a number of feature vectors for each node. 
The results of feature vectors are fed into a pooling layer, where we choose to average the different number of diverse vectors (i.e., 1-D average pooling) from the same midst as the final embedding vector. Such a process can be illustrated as Fig. \ref{fig:architecture} (Modeling Context Co-Occurrence) and \ref{fig:CNNvsCoANE} (bottom).
Finally, we formulate the process of the 1-D convolutional layer as well as the average pooling layer for node $v$, which can be respectively represented as below.
The convolutional part is given by:
$\mathbf{r}_{vij}^* = \sum \mathbf{R}_{vi} \odot \mathbf{\Theta}_j$,
for $i = 1,2,...,|context(v)|$ and $j = 1,2,...,d'$, where $\mathbf{r}_{vij}^* \in {\rm I\!R}$ is the $i$-th context's convolutional value from the $j$-th filter, $\mathbf{R}_{vi} \in {\rm I\!R^{c \times d}}$ is the $i$-th attribute-context matrix with midst node $v$, $\odot$ is Hadamard product, and  $\mathbf{\Theta}_j \in {\rm I\!R^{c \times d}}$ is the parameter matrix of the $j$-th filter.
The pooling part is given by:
$\mathbf{z}_{v} = \mathbf{\theta}_{pool}  \mathbf{R}_{v}^*$,
where $\mathbf{z}_{v}$ is the embedding vector of node $v$,  $ \mathbf{\theta}_{pool} \in {\rm I\!R^{1 \times |context(v)|}}$ is $1$-D average pooling operator, and $\mathbf{R}_{v}^* \in {\rm I\!R^{|context(v)| \times d'}}$ is the collection of convolution results from $\mathbf{r}_{vij}^*$, i.e., $\mathbf{R}_{v}^* = \{\mathbf{r}_{vij}^*\}$. 

\textbf{Discussion.}
The proposed method can be further discussed in comparisons of related studies. First, in Fig. \ref{fig:CNNvsCoANE}, we present an outline of CNN (top) and how CoANE utilizes the convolutional mechanism (bottom). Though Euclidean's input of CNN is not generally compatible with network data, 
we can use attribute-context matrices in a similar configuration as images so that CNN can be applied. 
CNN can recognize a cat picture via some filters to score and mine its tail and pupils. Similarly, CoANE can distinguish neighboring nodes via learning latent social circles, like ``baseball team'' and ``colleague'', using the various filters that learn higher weights on the correlated attributes like ``favorite sport'' and ``job.'' 

Second, compared to the state-of-the-arts GAE~\cite{Kipf-Welling:arXiv-2016} and VGAE~\cite{Kipf-Welling:arXiv-2016} that only uses fixed neighborhood in embedding learning, CoANE more emphasizes the learning in the specific patterns of similar attributes and positional information in the context. 
Hence, CoANE tends to incorporate more \text{precise} information from network structure and attributes, 
which is more capable of detecting various situations like one or multiple latent social circles (e.g., some friends likes baseball while another set of neighbors are colleagues who like jazz music).


\vspace{-4pt}
\subsection{Information Preservation and Objective Function}
While the network structure and node attributes have been encoded by the proposed attribute-context convolutional mechanism, now we present how to design the learning objective so that the derived embedding vectors can have structural and semantic preserving in the view of the context.  
The design of objective function can be divided into three parts: Positive Graph Likelihood, Contextually Negative Sampling, and Attribute Preservation. The first part is to preserve co-occurrence matrices $\mathbf{D}$ and $\mathbf{D^1}$ while the second is to make nodes tightly interconnected and overlapped with one another in terms of attributes that have similar embedding vectors. The third is to preserve node semantics by reconstructing the original attributes.
\subsubsection{Positive Graph Likelihood}
We take advantage of the idea of autoencoder that reconstructs features between input and output to impose the network structure into the embedding vectors. 
We extend the \textit{graph likelihood}~\cite{Abu-El-Haija-et-al:NIPS-2017} to reconstruct a new co-occurrence matrix $\mathbf{D}'$ by embedding vectors based on matrix factorization, and the aim is to minimize a reconstruction loss between $\mathbf{D}'$ and $\mathbf{D}$ (and $\mathbf{D^1}$). 
Let the embedding matrix be $\mathbf{Z}$, and $\mathbf{Z} = [\mathbf{L}|\mathbf{R}]$, where $\mathbf{L}, \mathbf{R} \in {\rm I\!R^{n \times \frac{d'}{2}}}$ are the left embedding and right embedding. The graph likelihood can be defined as follows.
\begin{equation} \label{eq:PO1}
    \displaystyle
    \prod_{v_i,v_j \in V}  \sigma (\mathbf{L}_i^T \mathbf{R}_j)^{\mathbf{D}_{ij}}
    \sigma (1-\mathbf{L}_i^T \mathbf{R}_j)^{I(\mathbf{E}_{ij} = 0)},
\end{equation}%
where $\sigma(x) = (1+\exp(-x))^{-1}$, $I$ is the indicator function, $L_i$ is the $i$-th row of $L$ and $R_j$ is the $j$-th row of $R$. 
The idea is to use embedding vectors to generate the co-occurrence matrix. An embedding learning can preserve more about the network if it can lead to high graph likelihood in the reconstruction of co-occurrence matrix $\mathbf{D}$.

However, the original graph likelihood needs to enumerate pairs of nodes (leading to high computation cost) and is not able to precisely emphasize on real edges of the network.
We leverage only the positively relational term and avoid sparser and higher cost negative term, and also strengthen the graph likelihood of one-hop nodes by adding $\mathbf{D^1}$. 

The adjusted graph likelihood $L_{pos}$ can be rewritten in the form of negative log-likelihood as below.
\begin{equation} \label{eq:PO}
    L_{pos} = -\sum_{v_i \in V}\sum_{v_j \in V, i\neq j} {\mathbf{\tilde{D}}_{ij}} log( \sigma (\mathbf{L}_i^T \mathbf{R}_j) )
\end{equation}%
where $\mathbf{\tilde{D}} = \mathbf{D^N} + \mathbf{D^1}$, and $\mathbf{D^N}$ is the normalized $\mathbf{D}$.

Note that the first-order (one-hop) neighbors should be more emphasized in learning node embeddings. This idea comes from random walk with restart (RWR)~\cite{rwr06}, i.e., personalized PageRank~\cite{page1999pagerank}. Although one-hop neighbors already have higher arriving probability values, the restarting probability in RWR is used to give one-hop neighbors much higher arriving probability values. The underlying intuition is that one-hop neighbors can directly represent the target node. Hence, we need to pay more attention to them by adding $\mathbf{D^1}$ to strengthen the graph likelihood. In addition, to further emphasize the contribution of one-hop neighbors in graph likelihood, we choose to use $\mathbf{D^N} + \mathbf{D^1}$, rather than the normalization of $\mathbf{D} + \mathbf{D^1}$. Adding $\mathbf{D^1}$ to the normalized $\mathbf{D}$ will give one-hop neighbors higher importance in learning node embeddings.

Furthermore, we think that the extremely small values in $\tilde{D}_{ij}$ could be noise when the graph is sparse. We want to lower down the effect of noisy structural information, and aim at preserving the most significant positive neighbors for each node. We consider the top-$k_p$ co-occurrence neighbors scored by $\tilde{D}_{ij}$. Specifically, for each node $v_i$, we compute $\mathbf{L}_i^T \mathbf{R}_j)$ for only $j \in$ top-$k_p({\mathbf{\tilde{D}}_{ij} | j = 1, ..., n})$. We determine the number of significant positive neighbors $k_p$ by $k_p = \underset{i= 1, ..., n}{max}(|context(v_i)|)$, where the $|context(v_i)|$ is the number of the sampled context nodes for $v_i$. We can treat $k_p$ as a kind of latent neighborhood size. With the preservation of top-$k_p$ significant neighbors, our loss function is able to not only pay more attention to preserve strong connections, but also better alleviate the impact of noisy neighborhood, which especially appears in sparse graphs that result in more less-frequent neighbors sampled by random walks.


\subsubsection{Contextually Negative Sampling}
In learning node embeddings, the basic idea is to make correlated nodes close in the embedding space. To better separate correlated and irrelevant nodes considering network structure and node attributes, we develop an efficient and effective \textit{contextually negative sampling} extended from negative sampling methods in word2vec~\cite{Mikolov-et-al:NIPS-2013} and AllVec~\cite{Xin-et-al:CL-2018}. In word2vec, the method is to select a small number of dissimilar samples 
and compute the loss using their logistic similarity by inner product. AllVec~\cite{Xin-et-al:CL-2018} considers all nodes out of the neighborhoods of the target word as the negative samples, and compute loss by square of inner product similarity. In addition to combining word2vec and AllVec, we further consider the contextual frequency of the target node and the negative samples so that the importance of negative samples can be modeled. 

We develop and expect our contextually negative sampling can push the embeddings of dissimilar nodes in both the network structure and node attributes to be away from each other according to the contextual frequency. Our negatively relational loss $L_{neg}$ for target node $v_i$ is given by: 

\begin{equation} \label{eq:NE}
    L_{neg}(v_i) =\sum_{j=1}^k E_{v_j \sim P_{ V^*(v_i)}}(a( \mathbf{z}_{v_i}^T  \mathbf{z}_{v_j})^2),
\end{equation}%
where $k$ is number of negative samples, the contextual noise distribution $P_{V}(v)$ is defined by  $\frac{|context(v)|}{\sum_{v \in V}|context(v)|}$ for any node set $V$,
$V^*(v) = \{v' \in V | v' \not\in context(v)\}$ is the set of nodes occurring out of the context of node $v$, and $a$ 
is a controlling constant deciding the strength of negative loss.

In detail, the negative loss $L_{neg}$ consists of two parts: the similarity between embedding vectors and the target-negative \textit{co-occurrence} probability for negative sampling. For the similarity, we utilize the square inner product similarity that is the same as existing NE methods and emphasizes the magnitude of similarity. Second, we aim at selecting the most significant negative samples because they need more power to be pushed away in the embedding space. 
We think the volume of a node's generated contexts is related to its main representation in the graph as well as a domination in a cluster. Therefore, we choose a node as an appropriate candidate for negative samples if its contexts cover the larger region (i.e., more contexts) in the network and have a lower correlation with the target node. That said, we sample nodes based on the contextual probability derived by its co-occurrence frequency of the contexts. The first $k$ nodes sampled from $V^*(v_i)$ are the more informative negative ones and considered as the negative samples for the target node $v_i$.
To deal with the high sampling cost when $V^*$ for each node has a diverse composition, we devise \textit{pre-sampling} and \textit{batch-sampling} to obtain negative nodes.
For the pre-sampling, first, we offline sample the nodes more than $k$, respecting to the contextual probability $P_{V}(v)$ as the negative sets. Then we select the first $k$ samples out of the contexts of target node $v_i$ as negative samples. 
For batch-sampling, we consider nodes in the batch for negative sampling during training. This can avoid too much probability computation and reduce the times of comparisons between nodes.  
The pre-sampling can reduce the training cost due to selecting negative targets before training. The batch-sampling only needs to handle nodes in each batch.


\subsubsection{Attribute Preservation}
In addition to encoding structural information, we also aim at preserving semantic information based on node features, i.e., attribute values, in the process of embedding learning. We utilize a multi-layer perceptron (MLP) decoder to reconstruct node features from the derived node embeddings. That said, we can obtain the reconstructed attributes $\hat{\mathbf{X}}_{v_i} = MLP(\mathbf{z}_{v_i})$, where 
MLP is constructed by stacking two hidden layers with the ReLU non-linear activation function.
The preservation loss is given as follows:
\begin{equation} \label{eq:AP}
    \displaystyle
    L_{att} = \gamma MSE(\hat{\mathbf{X}}, {\mathbf{X}}),
\end{equation}%
where MSE is mean square error, $\hat{\mathbf{X}}$ denote the reconstructed attribute values, $\gamma$ is the hyperparameter that controls the importance of the reconstruction effect. 


\subsubsection{Model Optimization}
The final objective function $L_{obj}$ of CoANE is given by Eq. (\ref{eq:OBJ}), which combines the positive graph likelihood Eq. (\ref{eq:PO}), the contextual negative sampling Eq. (\ref{eq:NE}), and the attribute preservation Eq. (\ref{eq:AP}). In the equation, the positive part ensures the preservation of structural and semantic proximities while the negative part pushes dissimilar ones away from each other and avoids overfitting.
\begin{equation} \label{eq:OBJ}
    \displaystyle
    L_{obj} = L_{pos} + \sum_{v_i \in V} L_{neg}(v_i) + L_{att}
\end{equation}%
We use batch gradient descent to optimize the parameters in $L_{obj}$ for the parameters of filters in our model, which nodes are randomly partitioned into several batches and make their corresponding embeddings update each epoch. 

The overall CoANE algorithm is outlined in Algorithm \ref{alg:CoANE}. After the pre-processing phase, we can obtain each node's contexts $context(v)$ and two matrices $\mathbf{D}$ and $\mathbf{D^1}$. Then we start the training phase and initialize both model parameters and embedding vectors of all nodes using Xavier uniformly initialization~\cite{glorot2010understanding}. At each training iteration, we apply batch updating, which further consists of Embedding Updating step and Loss Updating step, for $n_B$ nodes of $V_{n_B}$ sampled from $V$ without replacement. 
That is, we randomly split $V$ into a set of batches $V_B$ with size $n_B$ by the function $RandomlySplitBatch$, and perform updating procedures for each batch $V_{n_B}$ sampled from $V_B$.

For the sampled nodes, we first update their embeddings by following our convolutional process, and then the loss is calculated for updating model parameters. We expect that filters of the model are updated by contexts and are kept improved in generating embeddings. The whole updating process repeats until the convergence of $L_{obj}$ or the maximum step is achieved. Last, we need to refresh all embeddings the same as Embedding Updating step. 

\begin{algorithm}[!t]
\caption{CoANE algorithm}
\label{alg:CoANE}
\textbf{Input}: $\mathbf{G} =(V,\mathbf{E},\mathbf{X})$, repeat $r$, walk length $l$, scan probability $p(v)$, context length $c$, maximum epoch $N_{max}$, number of training nodes $n_B$ and number of negative samples $k$\\
\textbf{Parameter}: parameters of filters $\mathbf{\Theta}$\\
\textbf{Output}: node embedding $\mathbf{Z}$
\begin{algorithmic} 
\STATE{\#Pre-processing phase}
\FORALL{node $v$ in $V$} 
\STATE{$context(v)$ = RandomWalkProcess($V$, $\mathbf{E}$), $p(v)$, $c$, $r$, $l$)} 
\ENDFOR
\STATE Construct $\mathbf{D}$, $\mathbf{D^1}$ and negative samples $V_{neg}$ and initialize $\mathbf{Z}$ and $\mathbf{\Theta}$ 
\STATE{\#Training phase}
\FOR{$n = 1$ \TO $N_{max}$ }
\STATE{$V_{B} = $ $RandomlySplitBatch$($V$, $n_B$)
     \FOR{$V_{n_B}$ in $V_{B}$}
     \STATE{ 
             $\{ \mathbf{R}_{v} \}_{V_{n_B}}$ = AttributeConcat($\{ context(v) \}_{V_{n_B}}$, $X$)\\
             Update $\mathbf{Z}$ by $\{ \mathbf{z}_{v} \}_{V_{n_B}}$ = CNN($\mathbf{\Theta}$, $\{ \mathbf{R}_{v} \}_{V_{n_B}}$, $d'$)\\
             Compute $L_{pos}(\mathbf{Z}, \mathbf{D}, \mathbf{D^1})${ \# Sec. 3.3.1}\\
             Compute $L_{neg}(\mathbf{Z}, V_{neg}, k)$ \#{ Sec. 3.3.3}\\
             Compute $L_{att}(MLP(\mathbf{Z}), \mathbf{X})$ \#{ Sec. 3.3.4}\\
             $L_{obj}({V_{n_B}})= L_{pos} + L_{neg} + L_{att}$\\
             Update $\mathbf{\Theta}$ by GradientDescent($L_{obj}(V_{n_B})$)
            }
     \ENDFOR
}
\ENDFOR
\STATE{Renew $\mathbf{z}_{v}$ for all nodes $V$}\\
\RETURN{$\mathbf{Z}$}
\end{algorithmic}
\vspace{-3pt}
\end{algorithm}

\begin{table}[t]
\caption{Summary of the adopted datasets.}
\vspace{-5pt}
\label{tab:datasets}
\centering
\resizebox{1.0\linewidth}{!}{
\begin{tabular}{l|ccccc}  
\hline
Dataset	&	\#nodes	&	\#Attributes &	\#edges	& density  &	\#labels \\
\hline\hline
Cora	&	2708	&	1433	&	5278	&  0.0014  &	7	\\
\hline
Citeseer	&	3312	&	3703	&	4660	&  0.0008  &	6	\\
\hline
Pubmed	&	19717	&	500	&	44327	&  0.0002  &	3	\\
\hline
WebKB-Cornell	&	195	&	1703	&	286	&  0.0151  &	5	\\
WebKB-Texas	&	187	&	1703	&	298	&  0.0171  &	5	\\
WebKB-Washington	&	230	&	1703	&	417	&  0.0158  &	5	\\
WebKB-Wisconsin	&	265	&	1703	&	479	&  0.0137  &	5	\\
\hline
Flickr	&	7575	&	12047	&	239738	&  0.0084  &	9	\\
\hline
\end{tabular}
\vspace{-12pt}
}
\end{table}


\textbf{Complexity Analysis.} The time complexity relies on three parts: the convolutional mechanism with filters, the computation of co-occurrence matrices between contexts and attributes, and the attribute reconstruction for its preservation. 
The time complexity of the first part is $O(d*d'*c)$, where the context size $c$ is extremely smaller than the attribute dimension $d$ (i.e., $c\ll d$). This part is less than the state-of-the-art ANE model DANE ($O(d*d'*l$))~\cite{Gao-Huang:IJCAI-2018}, where $l$ is the number of parameters in the hidden layer and usually larger than embedding dimension $d'$. 
Second, for the co-occurrence matrices, since matrices are sparse, we use sparse data structure and operation. So, the complexity is $O(n*d')$, where $n = |V|$ is the number of nodes. Such complexity is also less than DANE that needs multi-proximity computation $O(n^2)$. We also leave the memory-efficient extensions via our batch updating. Then the complexity becomes $O(n_B*d')$ (i.e., $n_B\ll n$) for each sub-epoch.
Third, our attribute reconstruction is similar to DANE; however, the actual computation of DANE requires encoding, decoding, and multiple loss functions to capture structure and attribute information. Our CoANE only utilizes an attribute decoder for attribute preservation, which is a shallow structure, comparing to the deeper architecture of DANE.

%


\section{Experiments}

\subsection{Experiment Settings}

\textbf{Datasets.}
We employ five publicly available attributed network datasets, including Cora, Citeseer, WebKB, Pubmed~\footnote{Datasets for Cora, Citeseer, WebKB, and Pubmed available via \url{https://linqs.soe.ucsc.edu/data}}, and Flickr~\cite{huang2017label}
, for the experiments. The statistics of such five datasets are presented in Table \ref{tab:datasets}.
Each node is associated with a list of attributes and one class label. The class label is considered as the ground truth in the task of node label classification and node clustering. Note that since WebKB contains four small networks, we run the experiments separately, and report the average score.

\comment{
\begin{figure*}[t!]
  \centering
  \includegraphics[width=1.0\textwidth]{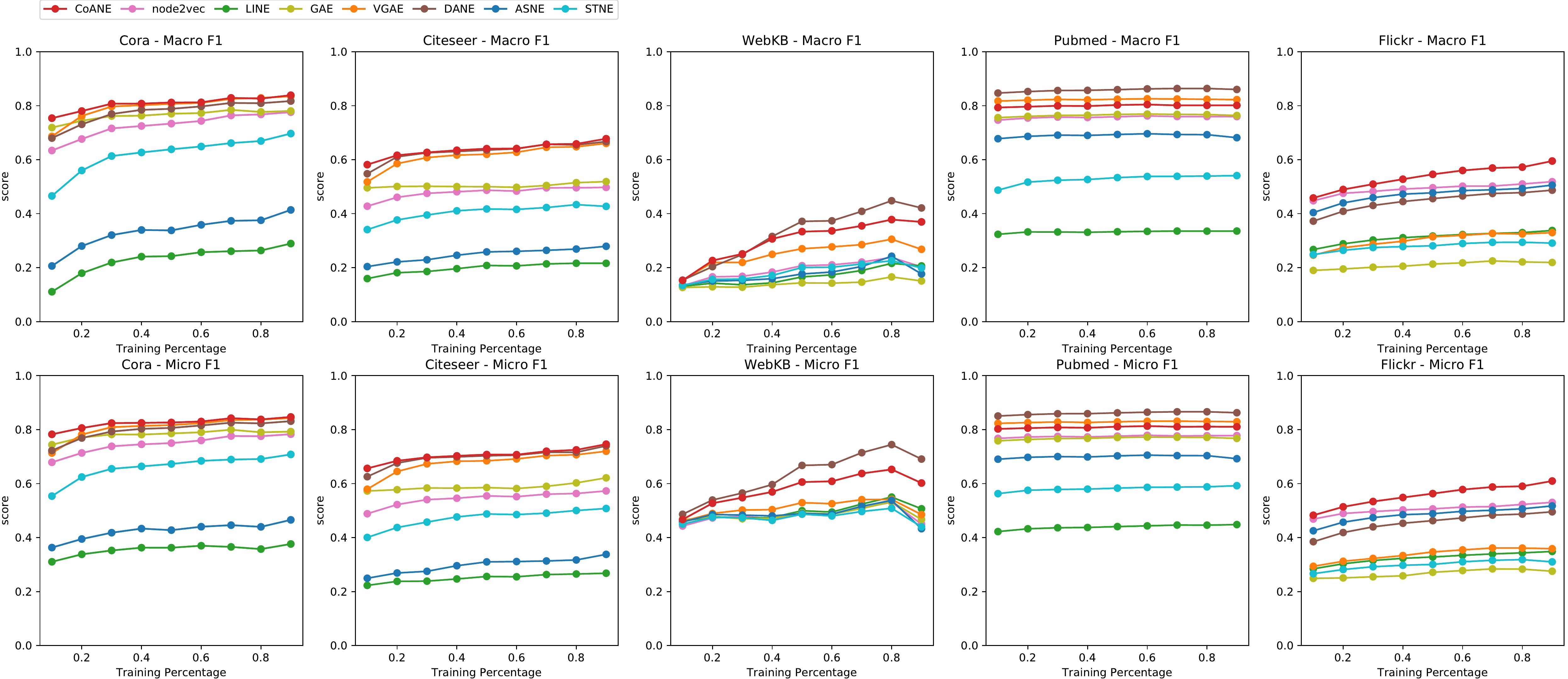}
  \caption{Results in Macro-F1 \& Micro-F1 for node classification by varying training percentage. Y-axis is the score of metric.
  }
  \label{fig:LC}
  \vspace{-6pt}
\end{figure*}
}

\begin{table*}[!t]
\caption{Macro and Micro F1 scores for node label classification of Cora, Citeseer, and Pubmed datasets. We mark the \textbf{rank-1} and \underline{rank-2} models in \textbf{bold} and \underline{underline}, respectively, in the following tables.}
\vspace{-5pt}
\label{tab:LC1}
\centering
\resizebox{1.0\linewidth}{!}{
\begin{tabular}{l|ccc|ccc |ccc|ccc |ccc|ccc}  
\hline
Dataset	&\multicolumn{6}{c|}{Cora}												&\multicolumn{6}{c|}{Citeseer}												&\multicolumn{6}{c}{Pubmed}												\\
\hline																																					
Training ratio	&5\%	&20\%	&50\%	&5\%	&20\%	&50\%							&5\%	&20\%	&50\%	&5\%	&20\%	&50\%							&5\%	&20\%	&50\%	&5\%	&20\%	&50\%\\
\hline
Method	&\multicolumn{3}{c|}{Macro F1} &\multicolumn{3}{c|}{Micro F1}&\multicolumn{3}{c|}{Macro F1} &\multicolumn{3}{c|}{Micro F1}&\multicolumn{3}{c|}{Macro F1} &\multicolumn{3}{c}{Micro F1}\\
\hline\hline
node2vec	&	0.663 	&	0.714 	&	0.750 	&	0.627 	&	0.677 	&	0.734 	&	0.437 	&	0.522 	&	0.555 	&	0.375 	&	0.461 	&	0.487 	&	0.760 	&	0.773 	&	0.776 	&	0.739 	&	0.754 	&	0.759 	\\
LINE	&	0.306 	&	0.338 	&	0.363 	&	0.093 	&	0.179 	&	0.243 	&	0.216 	&	0.238 	&	0.256 	&	0.115 	&	0.181 	&	0.208 	&	0.413 	&	0.433 	&	0.441 	&	0.319 	&	0.332 	&	0.333 	\\
GAE	&	\underline{0.737}	&	\underline{0.771}	&	0.786 	&	\underline{0.714} 	&	0.744 	&	0.770 	&	0.552 	&	0.577 	&	0.585 	&	0.471 	&	0.501 	&	0.500 	&	0.751 	&	0.764 	&	0.771 	&	0.749 	&	0.761 	&	0.768 	\\
VGAE	&	0.669 	&	0.782 	&	0.817 	&	0.649 	&	0.762 	&	0.807 	&	0.506 	&	0.645 	&	0.684 	&	0.441 	&	0.585 	&	0.620 	&	\underline{0.819}	&	\underline{0.826}	&	\underline{0.829}	&	\underline{0.812}	&	\underline{0.820}	&	\underline{0.824}	\\
GraphSAGE	&	0.622 	&	0.652 	&	0.657 	&	0.520 	&	0.565 	&	0.592 	&	0.608 	&	0.642 	&	0.653 	&	0.526 	&	0.567 	&	0.575 	&	0.645 	&	0.651 	&	0.654 	&	0.620 	&	0.625 	&	0.630 	\\
DANE	&	0.309 	&	0.366 	&	0.451 	&	0.086 	&	0.189 	&	0.316 	&	0.208 	&	0.281 	&	0.414 	&	0.057 	&	0.155 	&	0.294 	&	0.697 	&	0.759 	&	0.786 	&	0.701 	&	0.760 	&	0.787 	\\
ASNE	&	0.353 	&	0.395 	&	0.428 	&	0.178 	&	0.280 	&	0.338 	&	0.234 	&	0.269 	&	0.310 	&	0.155 	&	0.221 	&	0.258 	&	0.676 	&	0.697 	&	0.703 	&	0.663 	&	0.686 	&	0.693 	\\
STNE	&	0.488 	&	0.624 	&	0.673 	&	0.398 	&	0.560 	&	0.638 	&	0.319 	&	0.437 	&	0.488 	&	0.248 	&	0.377 	&	0.417 	&	0.546 	&	0.575 	&	0.583 	&	0.470 	&	0.517 	&	0.534 	\\
ARGA	&	0.477 	&	0.784 	&	0.808 	&	0.407 	&	0.761 	&	0.797 	&	0.312 	&	0.639 	&	0.675 	&	0.250 	&	0.583 	&	0.605 	&	0.407 	&	0.673 	&	0.680 	&	0.306 	&	0.678 	&	0.685 	\\
ARVGA	&	0.529 	&	0.808 	&	\underline{0.821}	&	{0.474}	&	\underline{0.783}	&	\underline{0.812}	&	0.341 	&	0.721 	&	0.736 	&	0.280 	&	0.647 	&	0.660 	&	0.400 	&	0.762 	&	0.781 	&	0.221 	&	0.754 	&	0.775 	\\
ANRL	&	0.673 	&	0.747 	&	0.758 	&	{0.622} 	&	0.709 	&	0.732 	&	\underline{0.696}	&	\underline{0.735}	&	\underline{0.746}	&	\underline{0.609}	&	\underline{0.679}	&	\underline{0.684}	&	0.707 	&	0.742 	&	0.759 	&	0.705 	&	0.742 	&	0.760 	\\\hline
CoANE	&	\textbf{0.767}	&	\textbf{0.818}	&	\textbf{0.840}	&	\textbf{0.737}	&	\textbf{0.787}	&	\textbf{0.824}	&	\textbf{0.723}	&	\textbf{0.744}	&	\textbf{0.759}	&	\textbf{0.628}	&	\textbf{0.680}	&	\textbf{0.696}	&	\textbf{0.825}	&	\textbf{0.842}	&	\textbf{0.851}	&	\textbf{0.816}	&	\textbf{0.836}	&	\textbf{0.847}	\\
\hline
\end{tabular}
\vspace{-5pt}
}
\end{table*}

\begin{table*}[!t]
\caption{Macro and Micro F1 scores for node label classification of WebKB, and Flickr datasets.}
\vspace{-5pt}
\label{tab:LC2}
\centering
\resizebox{.75\linewidth}{!}{
\begin{tabular}{l|ccc|ccc |ccc|ccc}  
\hline
Dataset	
&\multicolumn{6}{c|}{WebKB}												&\multicolumn{6}{c}{Flickr}												\\
\hline																																					
Training ratio	
&5\%	&20\%	&50\%	&5\%	&20\%	&50\%							
&5\%	&20\%	&50\%	&5\%	&20\%	&50\%				\\
\hline																																					
Method	&\multicolumn{3}{c|}{Macro F1} &\multicolumn{3}{c|}{Micro F1}&\multicolumn{3}{c|}{Macro F1} &\multicolumn{3}{c}{Micro F1}\\
\hline\hline	
node2vec	&	0.448	&	0.473	&	0.491	&	0.169	&	0.166	&	0.207	&	\underline{0.437}	&	\underline{0.489}	&	\underline{0.506}	&	\underline{0.400}	&	\underline{0.476}	&	\underline{0.496}	\\
LINE	&	0.455	&	0.478	&	0.500	&	0.142	&	0.143	&	0.166	&	0.257	&	0.303	&	0.328	&	0.236	&	0.288	&	0.317	\\
GAE	&	0.478	&	0.478	&	0.491	&	0.131	&	0.129	&	0.144	&	0.243	&	0.251	&	0.272	&	0.181	&	0.195	&	0.213	\\
VGAE	&	0.449	&	0.490	&	0.530	&	\underline{0.204}	&	0.220	&	0.270	&	0.287	&	0.312	&	0.347	&	0.234	&	0.274	&	0.314	\\
GraphSAGE	&	0.483	&	\underline{0.522}	&	0.563	&	0.183	&	0.202	&	0.254	&	0.145	&	0.158	&	0.170	&	0.098	&	0.123	&	0.142	\\
DANE	&	0.472	&	0.483	&	0.511	&	0.146	&	0.148	&	0.182	&	0.160	&	0.205	&	0.233	&	0.135	&	0.195	&	0.228	\\
ASNE	&	0.451	&	0.486	&	0.489	&	0.151	&	0.150	&	0.176	&	0.395	&	0.457	&	0.489	&	0.362	&	0.440	&	0.477	\\
STNE	&	0.432	&	0.476	&	0.487	&	0.169	&	0.156	&	0.200	&	0.251	&	0.282	&	0.301	&	0.222	&	0.264	&	0.281	\\
ARGA	&	0.434	&	0.483	&	0.528	&	0.152	&	0.192	&	0.254	&	0.155	&	0.189	&	0.213	&	0.131	&	0.168	&	0.201	\\
ARVGA	&	0.431	&	0.514	&	0.559	&	0.166	&	\underline{0.226}	&	0.286	&	0.159	&	0.109	&	0.128	&	0.095	&	0.022	&	0.043	\\
ANRL	&	\underline{0.494}	&	0.512	&	\underline{0.590}	&	0.198	&	0.190	&	\underline{0.310}	&	0.215	&	0.286	&	0.330	&	0.196	&	0.278	&	0.324	\\\hline
CoANE	&	\textbf{0.553}	&	\textbf{0.597}	&	\textbf{0.683}	&	\textbf{0.268}	&	\textbf{0.296}	&	\textbf{0.396}	&	\textbf{0.482}	&	\textbf{0.544}	&	\textbf{0.589}	&	\textbf{0.436}	&	\textbf{0.518}	&	\textbf{0.573}	\\
\hline
\end{tabular}
\vspace{-12pt}
}
\end{table*}

\textbf{Competing Methods.}
We consider well-known NE/ANE methods for performance comparison. 
The first two are commonly-used plain NE methods that consider no attributes: random-walk-based approaches node2vec~\cite{Grover-Leskovec:KDD-2016} (with parameters $p=q=1$) and LINE~\cite{Tang-et-al:TANG-2015} that preserves second-order graph proximity. 
The next three are state-of-the-art subgraph-aggregation-based ANE methods, including GAE/VGAE~\cite{Kipf-Welling:arXiv-2016} with $2$ layers (256-128) and GraphSAGE~\cite{hamilton2017inductive} with the mean aggregation. The following two are state-of-the-art graph-reconstruction-based ANE methods, including DANE~\cite{Gao-Huang:IJCAI-2018}~\footnote{To have fair comparison, we exclude the pre-training part in DANE source code because the pre-training part is never mentioned in the paper and all of the competing methods (including our CoANE) do not consider pre-training. That said, we faithfully utilize the end-to-end training of DANE as one of our competing methods.
} with $2$ layers (128-64), and ASNE~\cite{Liao-et-al:IEEE-2018}. 
For the context-modeling-based approach, we consider STNE~\cite{liu2018content}, in which the layer sizes of encoder and decoder are all $64$~\footnote{All settings of encoder and decoder and the generation of contexts follow the original STNE paper. 
}. Our CoANE is also a context-based approach like STNE. 
Last, CoANE is further compared with two recent state-of-the-art approaches ANRL~\cite{zhang2018anrl} and ARGA/ARVGA~\cite{pan2018adversarially} with layers (256-128) and discriminator (128-512) that follow the settings mentioned in the original paper.
For node2vec, DANE, and ANRL, we set the parameters for the random walk: the window size $=10$, the walk length $=80$, and the number of walks $r = 10$. 
To have a fair comparison in generating low-dimensional node embeddings, the dimension of node embedding vector is set $d'= 128$ for all methods. 
For CoANE, we set the number of walks $r=1$, the walk length $=80$, the scan parameter $t = 10^{-5}$, and the number of negative samples $k = 20$. 
Besides, we consider a 2-layer multi-layer perceptron $MLP$ with ReLU non-linear mapping for attribute preservation, and
we choose Adam optimizer with learning rate $=0.001$~\cite{kingma-Ba:arXiv-2014}. 

On the tuning of CoANE hyperparameters, the negative loss controller $a$ in Eq. (\ref{eq:NE}), the context window size $c$, and the attribute preservation controller $\gamma$ in Eq. (\ref{eq:AP}) are tuned by the validation set with ranges: $a \in [\text{1e-5, 1e-1}]$, $c \in [3, 5, 7, 9, 11]$ and $\gamma \in [\text{1e3, 1e7}]$, respectively. Besides, we apply pre-sampling to obtain negative samples in Eq.~(\ref{eq:PO}) and Eq.~(\ref{eq:NE}) for the denser graphs (i.e., WebKB and Flickr), and apply batch-sampling in Eq.~(\ref{eq:PO}) and Eq.~(\ref{eq:NE}) for the sparser graphs (i.e., Cora, Citeseer, and Pubmed).
\vspace{-6pt}

%

\subsection{Main Results}

\textbf{Node Label Classification.}
For node classification, we randomly select training and testing samples by varying the percentage of the training set in $5$\%, $20$\%, and $50$\%, and employ one-vs-rest logistic regression classifier with L2 regularization (by following the common-used settings~\cite{Grover-Leskovec:KDD-2016}). We utilize Macro-F1 and Micro-F1 as the evaluation metrics, in which higher scores indicate better performance. 
The results are exhibited in Table.~\ref{tab:LC1}, and ~\ref{tab:LC2}. 
It can be clearly observed that CoANE consistently leads to the best performance among $11$ competing methods across five datasets and two metrics. The improvement of CoANE also keeps stable with various training percentages. Such results prove the usefulness of modeling context co-occurrence (i.e., latent social circles), which is not considered in baselines. Typical models node2vec, ASNE, LINE, and STNE cannot produce higher scores as they cannot well depict mutually-correlated nodes by distinguishing the contribution of links. The autoencoder in DANE is hard to capture the correlation between node attributes and graph structure, so it cannot keep competing scores. Besides, GAE, VGAE, GraphSAGE, and ANRL consider neighborhood aggregation that preserves the local connectivity surrounded each node. Therefore, their scores are higher than other methods, and are closer to our CoANE. However, the discriminator adopted by ARGA and ARVGA built upon GAE/VGAE-based models can improve the performance, but make scores a bit unstable. These outstanding performances of CoANE also demonstrates the effectiveness of its three-way objective.

\begin{table*}[!t]
\caption{AUC scores for link prediction (Left); NMI scores for node clustering (Right).}
\vspace{-5pt}
\label{tab:LP_CLU}
\centering
\resizebox{0.85\linewidth}{!}{
\begin{tabular}{l|ccccc ||ccccc}  
\hline
Task	
&\multicolumn{5}{c||}{Link Prediction}
&\multicolumn{5}{c}{Node Clustering}\\
\hline					
Method$\backslash$Dataset	&	Cora	&	Citeseer	&	Pubmed			&	WebKB	&	Flickr							&	Cora	&	Citeseer	&	Pubmed			&	WebKB	&	Flickr							\\
\hline\hline																																					
node2vec	&	0.896	&	0.901	&	0.927			&	0.684	&	0.748							&	0.367	&	0.149	&	0.273			&	0.058	&	\underline{0.165}							\\
LINE	&	0.632	&	0.626	&	0.754			&	0.664	&	0.648							&	0.052	&	0.005	&	0.003			&	0.074	&	0.088							\\
GAE	&	0.921	&	0.934	&	0.947			&	0.507	&	0.903							&	0.374	&	0.198	&	0.228			&	0.007	&	0.109							\\
VGAE	&	0.923	&	0.949	&	\textbf{0.975}			&	0.639	&	0.914							&	0.361	&	0.157	&	\underline{0.275}			&	0.092	&	0.131							\\
GraphSAGE	&	0.757	&	0.836	&	0.744			&	0.700	&	0.502							&	0.382	&	0.305	&	0.147			&	0.128	&	0.037							\\
DANE	&	0.663	&	0.768	&	0.869			&	0.635	&	0.901							&	0.021	&	0.032	&	0.148			&	0.083	&	0.015							\\
ASNE	&	0.571	&	0.586	&	0.792			&	0.448	&	0.848							&	0.073	&	0.005	&	0.165			&	0.078	&	0.111							\\
STNE	&	0.846	&	0.885	&	0.880			&	0.670	&	0.913							&	0.207	&	0.068	&	0.038			&	0.069	&	0.081							\\
ARGA	&	\underline{0.941}	&	0.966	&	0.920			&	0.614	&	\underline{0.925}							&	0.452	&	0.181	&	0.211			&	0.092	&	0.066							\\
ARVGA	&	0.927	&	\underline{0.972}	&	0.877			&	\underline{0.765}	&	\textbf{0.926}							&	\underline{0.530}	&	0.381	&	0.244			&	0.104	&	0.108							\\
ANRL	&	0.871	&	0.965	&	0.769			&	0.752	&	0.601							&	0.391	&	\underline{0.407}	&	0.099			&	\underline{0.132}	&	0.014							\\\hline
CoANE	&	\textbf{0.947}	&	\textbf{0.982}	&	\underline{0.969}			&	\textbf{0.784}	&	\textbf{0.926}							&	\textbf{0.544}	&	\textbf{0.435}	&	\textbf{0.313}			&	\textbf{0.180}	&	\textbf{0.211}\\\hline							
\end{tabular}
\vspace{-9pt}
}
\end{table*}

\begin{table}[!t]
\caption{NMI for Clustering on WebKB networks.}
\vspace{-5pt}
\label{tab:CLU_w}
\centering
\begin{tabular}{l|cccc}  
\hline
Method$\backslash$Dataset	&	Cornell	&	Texas	&	Washington	&	Wisconsin\\
\hline													
node2vec	&	0.066 	&	0.070 	&	0.044 	&	0.053 					\\
LINE	&	0.066 	&	0.093 	&	0.085 	&	0.051 					\\
GAE	&	0.002 	&	0.000 	&	0.027 	&	0.000 					\\
VGAE	&	0.086 	&	0.081 	&	0.103 	&	0.096 					\\
GraphSAGE	&	\underline{0.105}	&	\underline{0.157}	&	0.140 	&	0.111 					\\
DANE	&	0.067 	&	0.087 	&	0.118 	&	0.061 					\\
ASNE	&	0.066 	&	0.094 	&	0.103 	&	0.047 					\\
STNE	&	0.071 	&	0.088 	&	0.065 	&	0.052 					\\
ARGA	&	0.086 	&	0.093 	&	0.099 	&	0.091 					\\
ARVGA	&	0.091 	&	0.094 	&	0.128 	&	0.101 					\\
ANRL	&	0.114 	&	0.116 	&	\underline{0.167}	&	\underline{0.131}					\\\hline
CoANE	&	\textbf{0.191}	&	\textbf{0.200}	&	\textbf{0.181}	&	\textbf{0.148}					\\
\hline
\end{tabular}
\vspace{-8pt}
\end{table}

\textbf{Node Clustering.} 
We also conduct experiments to examine whether the embeddings generated by CoANE can produce effective node clustering.
We employ K-means algorithm, along with the embedding vectors of nodes, to perform clustering. The number $K$ of clusters is given by the number of ground-truth labels. The normalized mutual information (NMI) is used as the evaluation metric. Higher NMI scores imply better performance. Nodes with the same labels are treated as clusters. 
The results are summarized in Table~\ref{tab:LP_CLU} (right). Additional results on four networks in WebKB data are presented in Table~\ref{tab:CLU_w}.
We can apparently find that the proposed CoANE significantly outperforms state-of-the-art methods across five datasets and four networks for WebKB dataset.
We can find that the competing methods have lower and unstable scores in all datasets. 
It implies that 
modeling attribute-context matrices of specific contexts can better depict the representational features for the target node, which cannot be well captured by node2vec, GAE, VGAE, GraphSAGE, and ANRL. Besides, LINE and ASNE that model lower-order neighbors are hard to distill deeper relationships among nodes. STNE only preserves local features and is hard to encode high-order neighborhoods that can be organized as latent social circles. Moreover, although ARGA and ARVGA further train the discriminators to enhance the preservation of graph topology, their models cannot well capture the meaningful neighborhood that is composed by both graph structure and node features.

\textbf{Link prediction.}
For each dataset, we randomly choose $ 70$\%, $10$\%, and $20$\% edges as the training, validation, and testing sets, respectively.
While training the link prediction model, we randomly sample the same number of non-existing links as negative instances, and ensure the negative instances are not replicated in both sets. Then we employ logistic regression as the classifier. We follow the settings of node2vec~\cite{Grover-Leskovec:KDD-2016}: using the Hadamard product of embedding vectors to generate the feature vector of each node pair. The area under the ROC curve (AUC) is utilized as the evaluation metric. 
The results are shown in Table~\ref{tab:LP_CLU} (left). . 
We can find that the embeddings generated by CoANE outperform those generated by most of the competing methods across five datasets. Such promising results exhibit the effectiveness of modeling a more precise attribute-context correlation by the proposed context generator and convolutional mechanism. That is, competing methods via lower-order proximities (i.e., LINE and ASNE) are obviously not able to distinguish loosely-connected nodes. node2vec cannot incorporate node attributes into embedding learning. 
Besides, STNE preserves the first-order neighbors and requires more dimensions to encode higher-order latent correlation between nodes, which results in worse performance.
Since GAE and VGAE have better information aggregation between network structure and node attributes, they produce higher scores than GraphSAGE and DANE. The discriminator in ARGA/ARVGA and the joint structure-attribute modeling in ANRL make them have relatively better performance. 
Nevertheless, without the learning of latent social circles, all competing methods cannot achieve stable performance in all datasets. Our CoANE can produce higher NMI scores in most of the five datasets.

\begin{figure}[!t]
  \centering
  \includegraphics[width=0.85\linewidth]{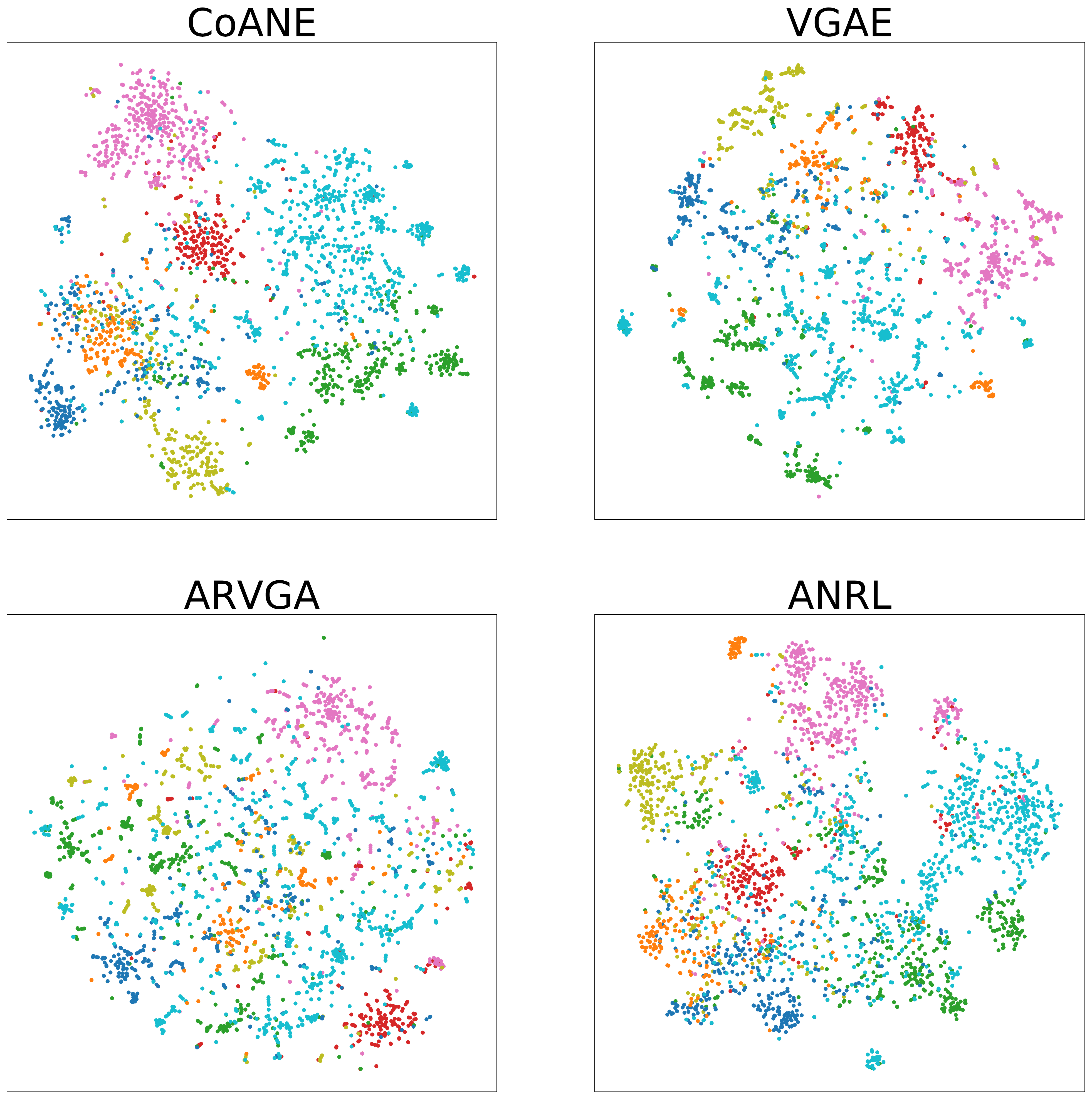}
  \vspace{-8pt}
  \caption{Visualizing various approaches for Cora data.}
  \vspace{-8pt}
  \label{fig:VP}
\end{figure}

\textbf{Summary.}
In summary, our CoANE can mostly outperform eleven competing methods on three tasks across five datasets, i.e., CoANE performs the best in 39 out of 40 cases\footnote{40 cases = 5 datasets $\times$ 8 task settings, in which 8 tasks consists of 3 training percentages (5\%, 20\%, 50\%) for node label classification with 2 metrics (Macro F1 and Micro F1), node clustering (NMI), and link prediction (AUC).}. In the only one case (1 out of 40 cases) that CoANE cannot work the best (i.e., the link prediction task on Pubmed dataset)
, CoANE's performance is still quite close to the best competing methods. We find that most competitive models have the neighborhood aggregation mechanism (e.g., VGAE and ARVGA) because generating embeddings by aggregating information from neighbors can better fuse graph structure and node attributes. The superiority of our CoANE comes from not only neighborhood aggregation (by the convolutional mechanism), but the finer-grained modeling of neighborhoods in terms of context co-occurrences, i.e., the latent social circles. In other words, the merit of CoANE mainly lies in the convolutional mechanism that captures both intra-hop and inter-hop feature correlation, and the comprehensive design of loss function to better preserve structural and semantic knowledge in node embeddings. In addition, CoANE leads to the best time efficiency while the strong baselines VGAE and ARGA require more training time. Last, we have realized that CoANE cannot work the best for predicting links in the graph with extremely high sparsity (i.e., Pubmed whose density is $0.002$). We think the reason is that high sparsity makes the latent social circles less evidential during training.

\begin{figure*}[!t]
     \centering
     \begin{subfigure}[b]{0.24\textwidth}
         \centering
         \includegraphics[width=\textwidth]{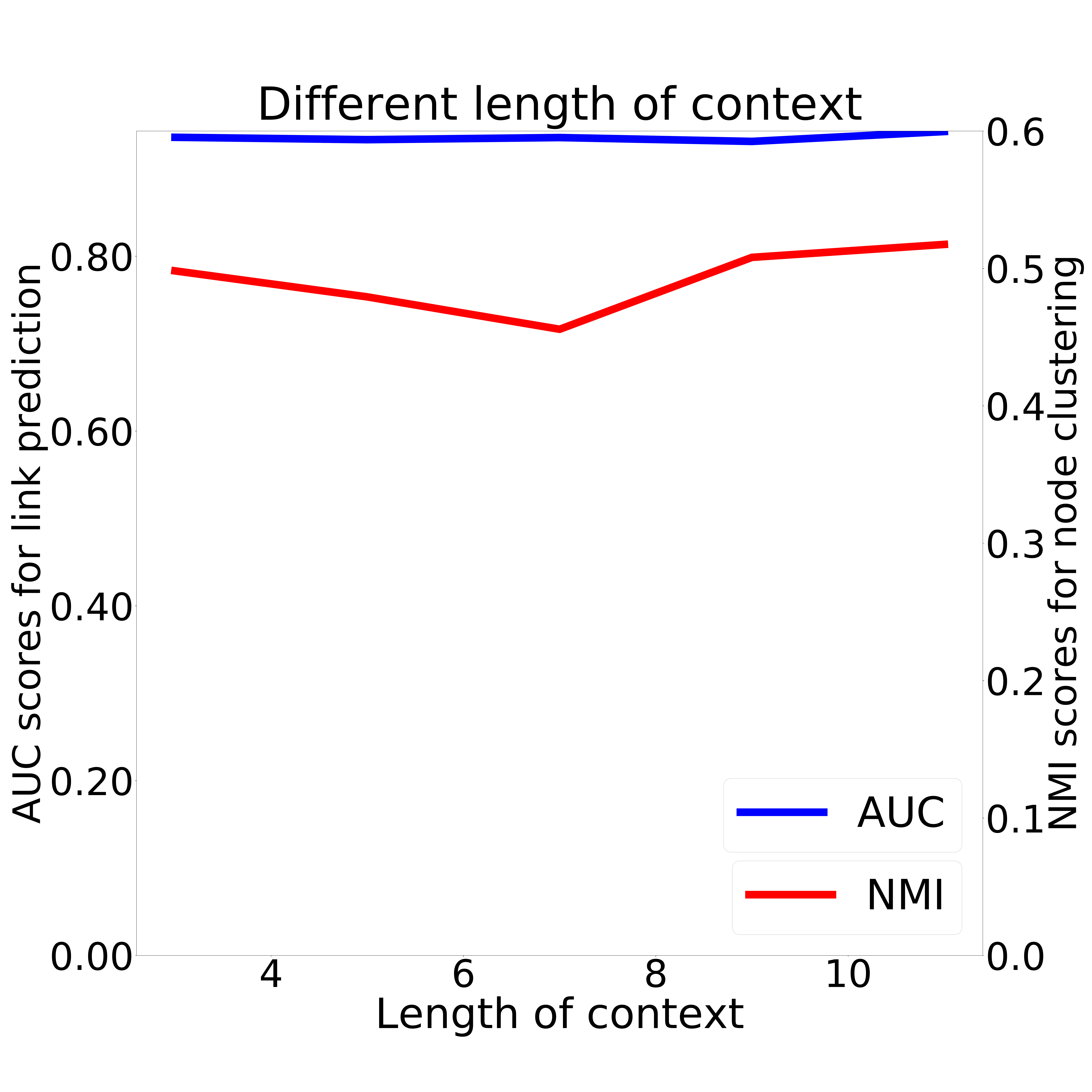}
         \caption{context length}
         \label{fig:P_C}
     \end{subfigure}
     \hfill
     \begin{subfigure}[b]{0.24\textwidth}
         \centering
         \includegraphics[width=\textwidth]{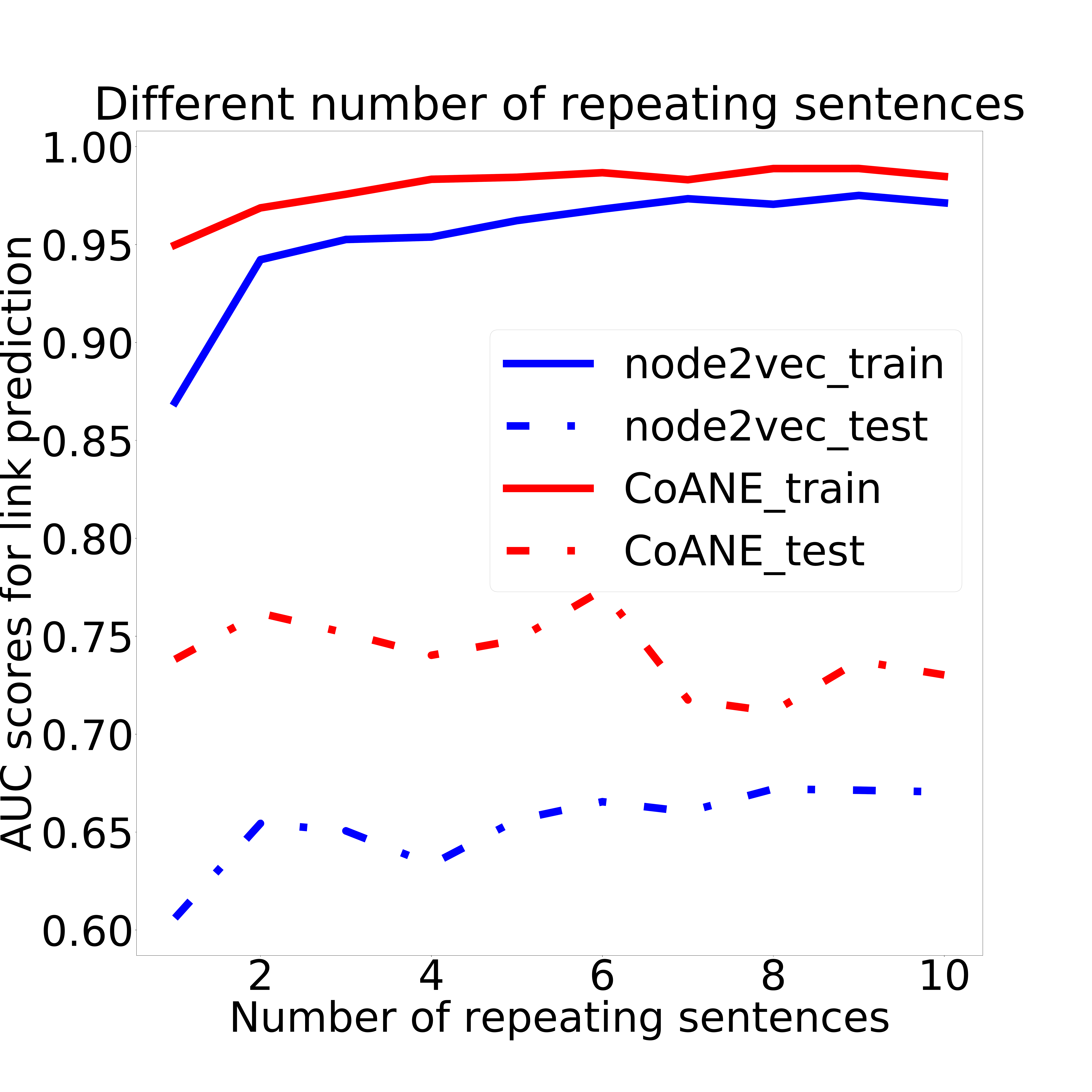}
         \caption{\#sampling walks}
         \label{fig:P_R}
     \end{subfigure}
     \hfill
     \begin{subfigure}[b]{0.24\textwidth}
         \centering
         \includegraphics[width=\textwidth]{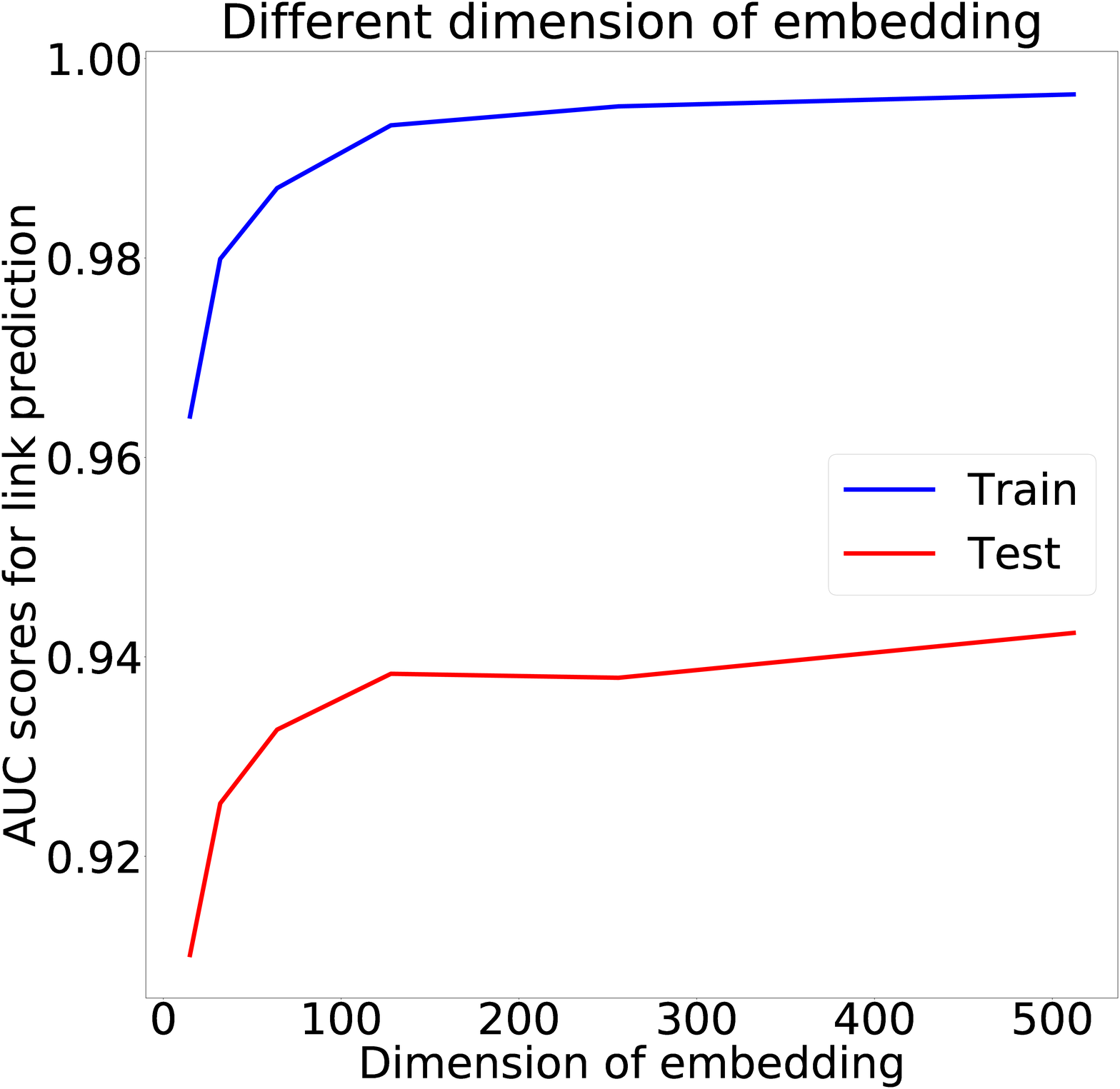}
         \caption{embedding dimension}
         \label{fig:P_d}
     \end{subfigure}
     \hfill
     \begin{subfigure}[b]{0.24\textwidth}
         \centering
         \includegraphics[width=\textwidth]{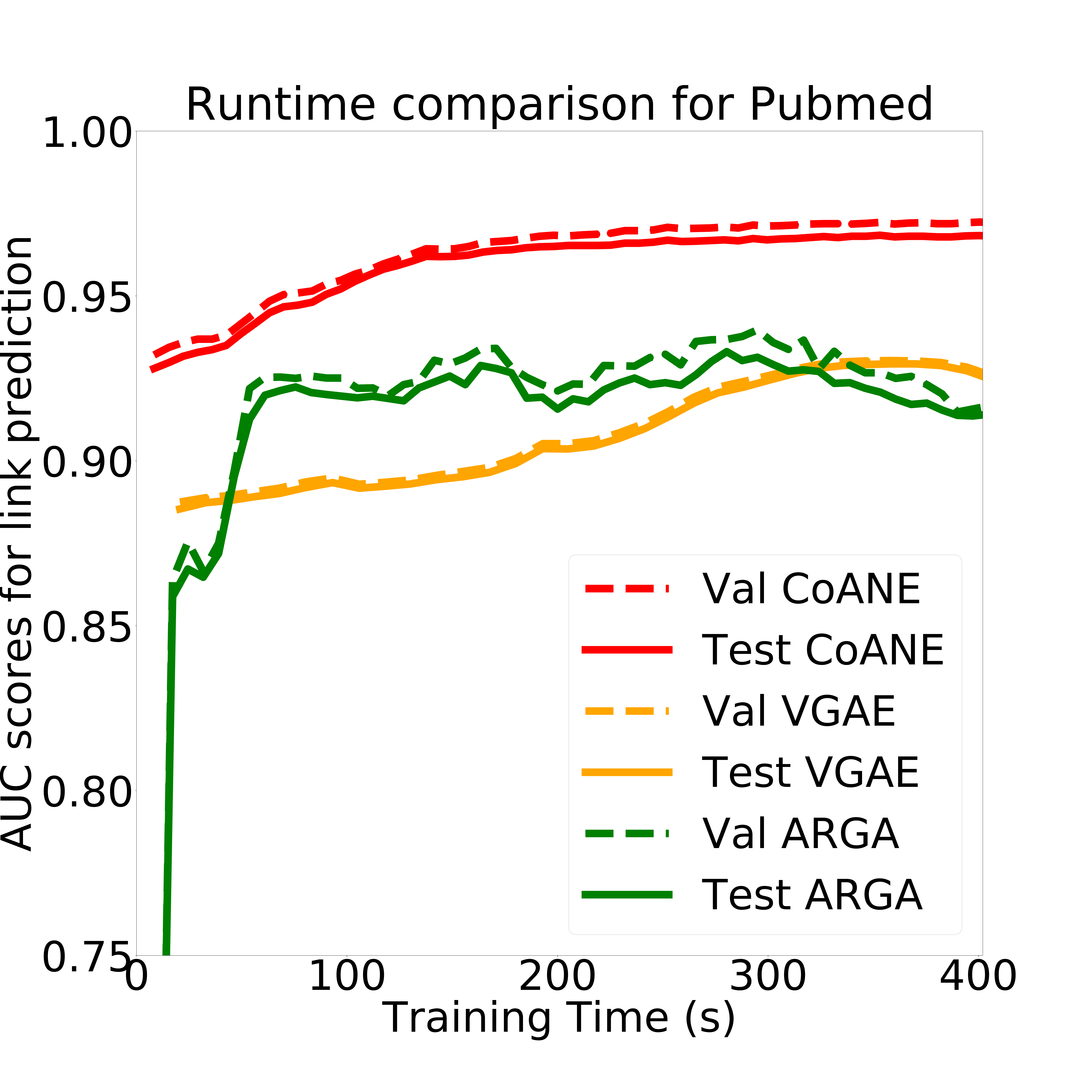}
         \caption{runtime in seconds}
         \label{fig:RT}
     \end{subfigure}
        \caption{(a)-(c) Sensitivity analysis for (a) length of contexts, (b) number of sampled sequences for random walk, (c) embedding dimension. (d) Runtime analysis.}
        \label{fig-sen}
        \vspace{-2pt}
\end{figure*}

\subsection{Model Analysis}
\textbf{Embedding Visualization.} 
To further present the embedding results of the proposed CoANE, we visualize the node representation using t-SNE~\cite{tsne2008}. The plots allow us to see the global property of embeddings and understand whether nodes with the same labels are close enough in the embedding space. Note that due to the page limit, we only show the results of three competing methods using Cora dataset. The visualization is shown in Fig.~\ref{fig:VP}, in which nodes are colored according to their labels.
It can be apparently found that our CoANE can lead to more compact and well-separated clusters than VAGE and ARVGA. Although ARNL can also generate a clear separation of groups, our CoANE can further attract those possessing the same labels to be much closer and push those with different labels farther away from each other. 
Such visualization also unveils why CoANE can achieve better performance on both supervised and unsupervised tasks.

\textbf{Sensitivity Analysis.}
We analyze three hyperparameters and training runtime in CoANE: (a) length of contexts, (b) number of sampled sequences for random walk, and (c) embedding dimension. 
For (a) and (b), we use WebKB and compare with node2vec, and consider CoANE without the attribute preservation for the analysis. For (c), we apply link prediction for CoANE in various embedding dimensionalies, and present the results of training and test sets. 

(a) \textit{Length of contexts.} The length of contexts represents the size of neighbors. Larger sizes would result in high computation cost.
We show how the context length affects the performance of link prediction (AUC) and node clustering (NMI) in Fig.~\ref{fig:P_C}. 
The stable results in both AUC and NMI inform us that as the size of neighbors gets higher (i.e., larger context length), the performance can be kept and does not change too much. The context length $=3$ may be enough. We think the reason is that local information is enough for link prediction and node clustering. 

(b) \textit{Number of sampled sequences.} In the random walk, the number of sampled node sequences can affect the quality of embeddings because fewer sampled sequences provide less neighborhood information. 
By varying the number of sampled sequences, as shown in Fig.~\ref{fig:P_R}, we compare the performance of link prediction (AUC) between node2vec and CoANE using WebKB data.
We can find that node2vec needs at least two sampled sequences for achieving stable performance. As for CoANE, requiring only one sampled sequence can lead to earlier stable results. 

(c) \textit{Dimension of embeddings.} We show how the dimension of embeddings affects the performance of link prediction. 
The results shown in Fig.~\ref{fig:P_d} demonstrate that a bit higher dimensionality can lead to better performance in both training and testing. The performance gets stable when the dimensionality is larger than $150$, implying that most information on network structure and node attributes is preserved. 

\textbf{Runtime Analysis.}
We present runtime analysis to understand the time efficiency of competing methods using the larger-scale Pubmed data. We conduct link prediction and report the AUC scores (y-axis) in validation and testing, and the training time in seconds for each epoch (x-axis) for CoANE and two stronger competing methods, VGAE and ARGA. The experiment is performed under Google Cloud Platform whose computing environment contains 8 vCPU with 30 GB memory and 1 NVIDIA Tesla K80 (12 GB GDDR5). The results are exhibited in Fig.~\ref{fig:RT}. We can clearly find that both VGAE and ARGA require more training time to reach their converged performance. 
It is because VGAE needs to indirectly aggregate the higher-order context neighbors, which results in taking more time to capture latent features. With the benefit of the discriminator, ARGA gains the improvement of effectiveness and efficiency. However, ARGA ignores the aggregation of related neighbors and cannot find all latent links. CoANE leads to high AUC scores with fast convergence (in only one epoch) in terms of training time. It is because CoANE can extract more representative contexts and utilize multiple filters that efficiently can learn important features.

\begin{figure}[!t]
     \centering
     \begin{subfigure}[b]{0.2415\textwidth}
         \includegraphics[width=\textwidth]{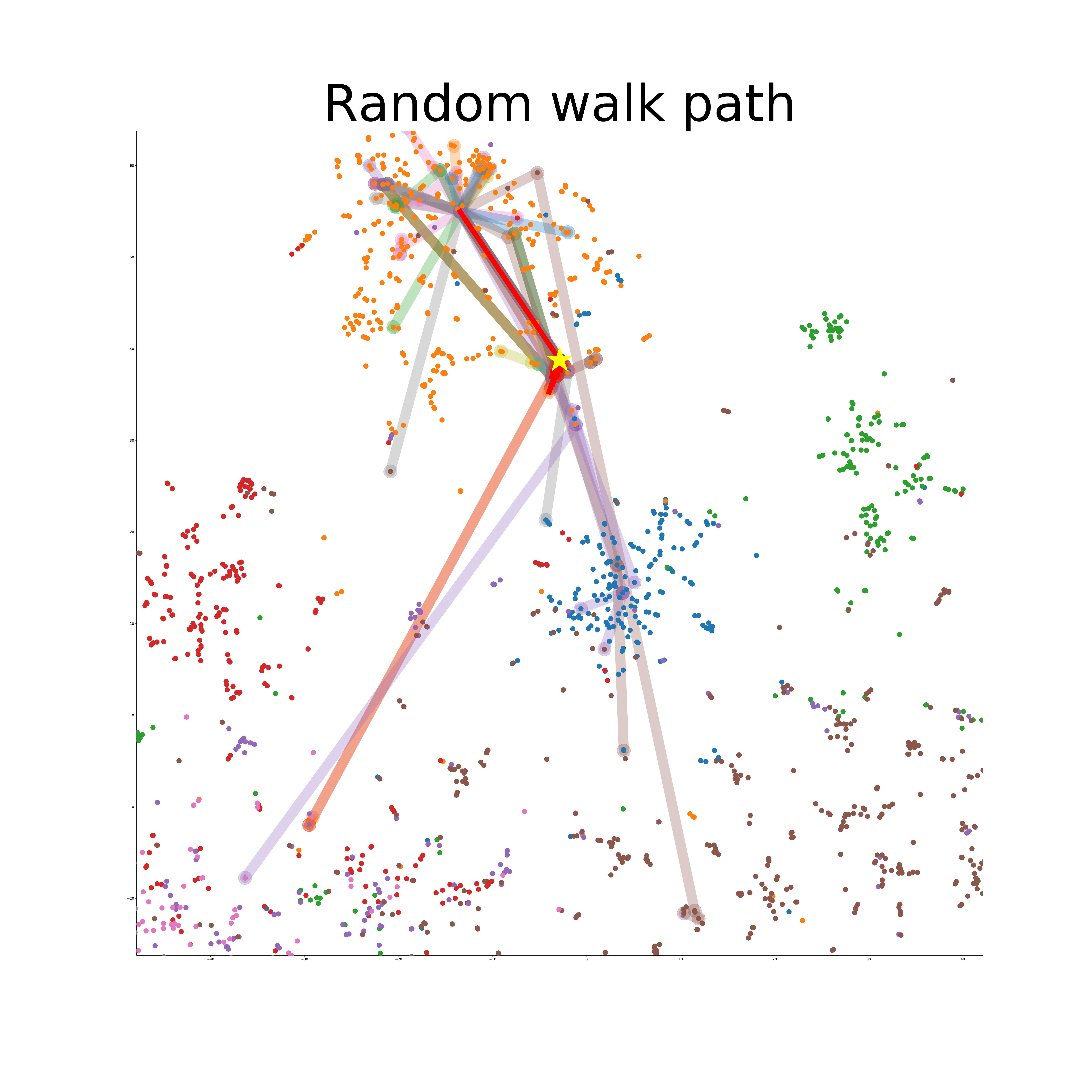}
         \caption{Random walk paths}
         \label{fig:Pre_rw1}
     \end{subfigure}
     \hfill
     \begin{subfigure}[b]{0.2415\textwidth}
         \includegraphics[width=\textwidth]{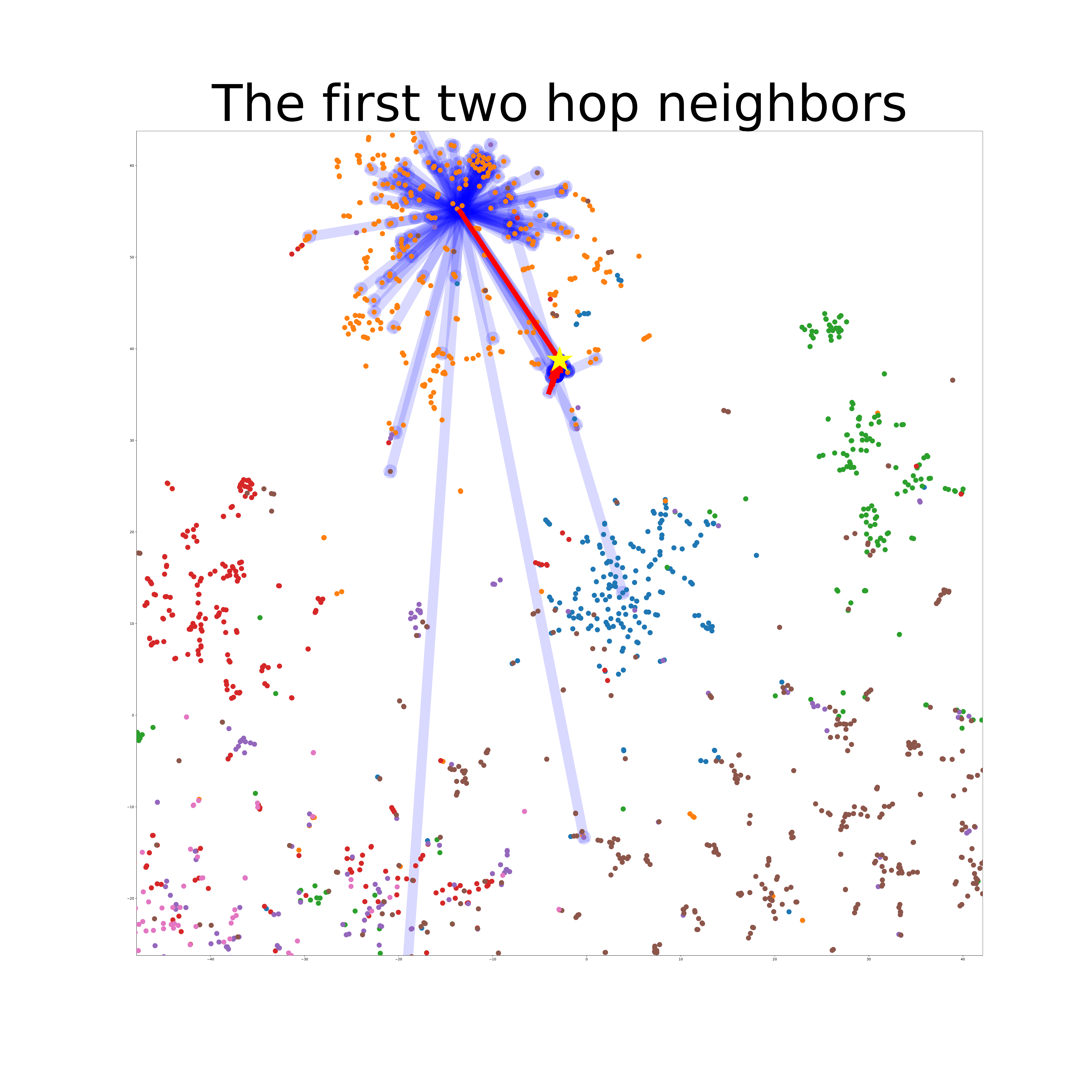}
         \caption{First two hop neighbors}
         \label{fig:Pre_rw2}
     \end{subfigure}
        \caption{Analyzing the neighbor selection via Cora dataset.}
        \label{fig-pre1}
        \vspace{-14pt}
\end{figure}

\begin{figure*}[t]
     \centering
     \begin{subfigure}[b]{0.24\textwidth}
         \centering
         \includegraphics[width=\textwidth]{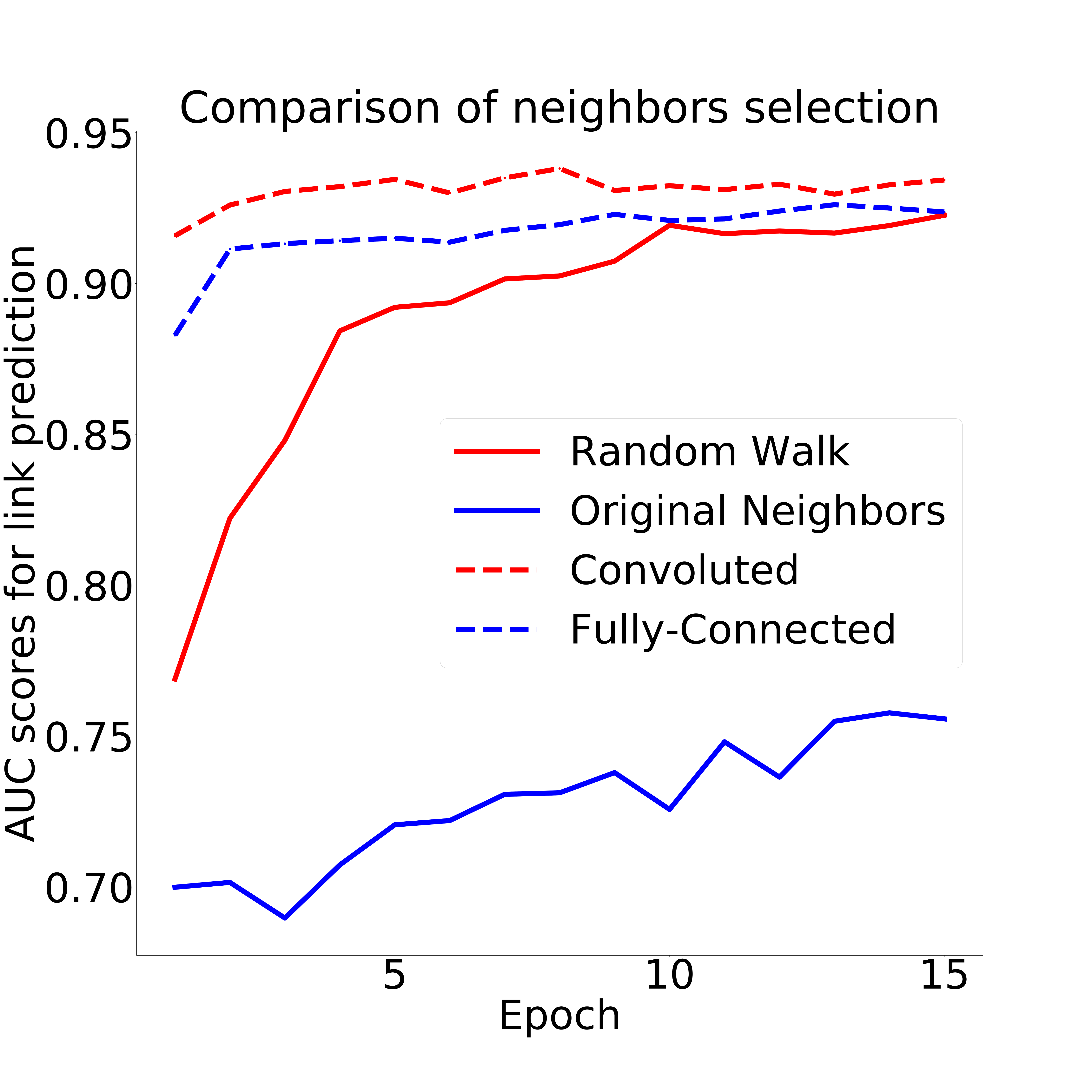}
        \caption{Neighbor and extractor.}
        \vspace{-2.5pt}
         \label{fig:P_rw1_CNN}
     \end{subfigure}
     \hfill
     \begin{subfigure}[b]{0.26\textwidth}
         \centering
         \includegraphics[width=\textwidth]{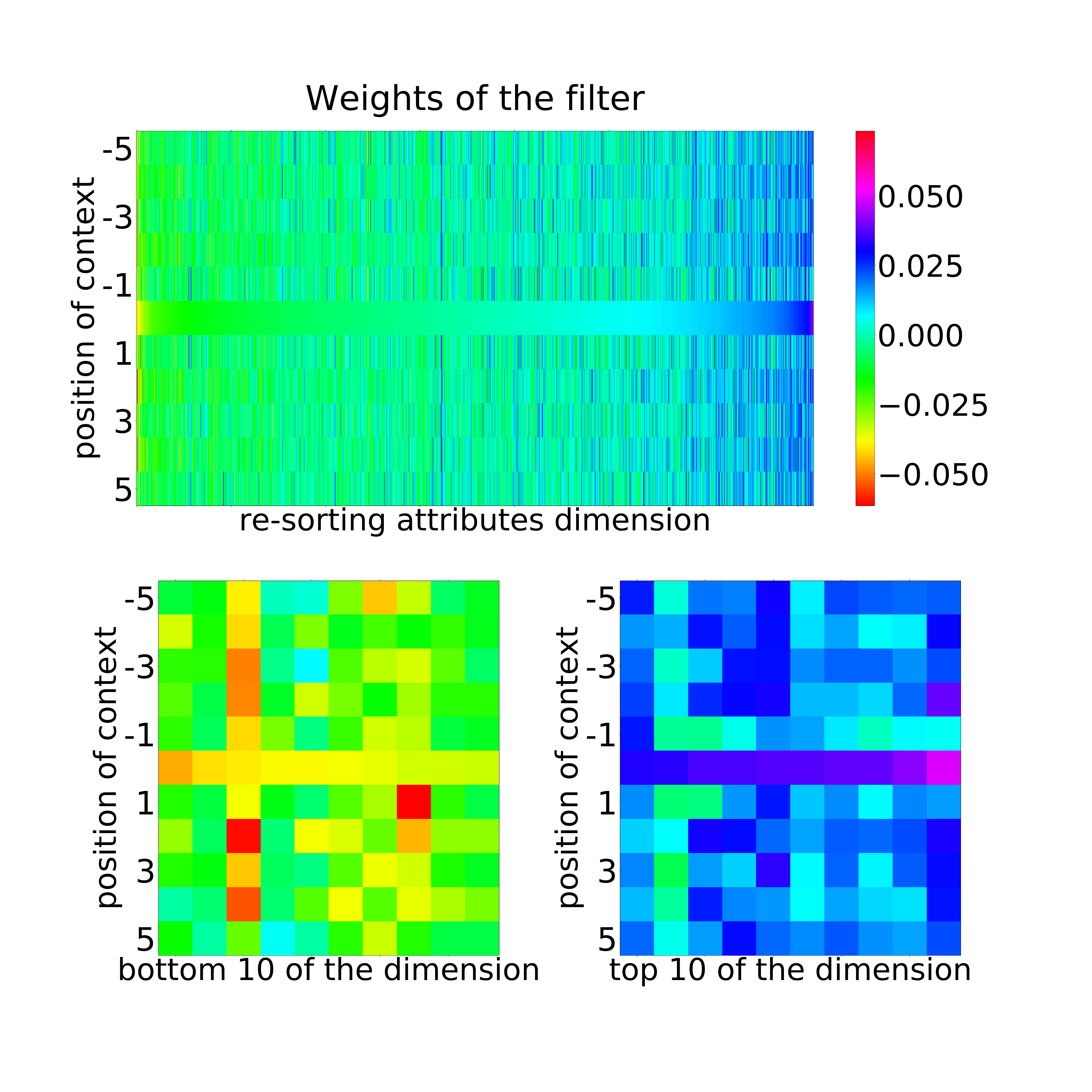}
         \caption{Visualizing filters.}
         \vspace{-2.5pt}
         \label{fig:P_F}
     \end{subfigure}
     \hfill
     \begin{subfigure}[b]{0.24\textwidth}
         \centering
         \includegraphics[width=\textwidth]{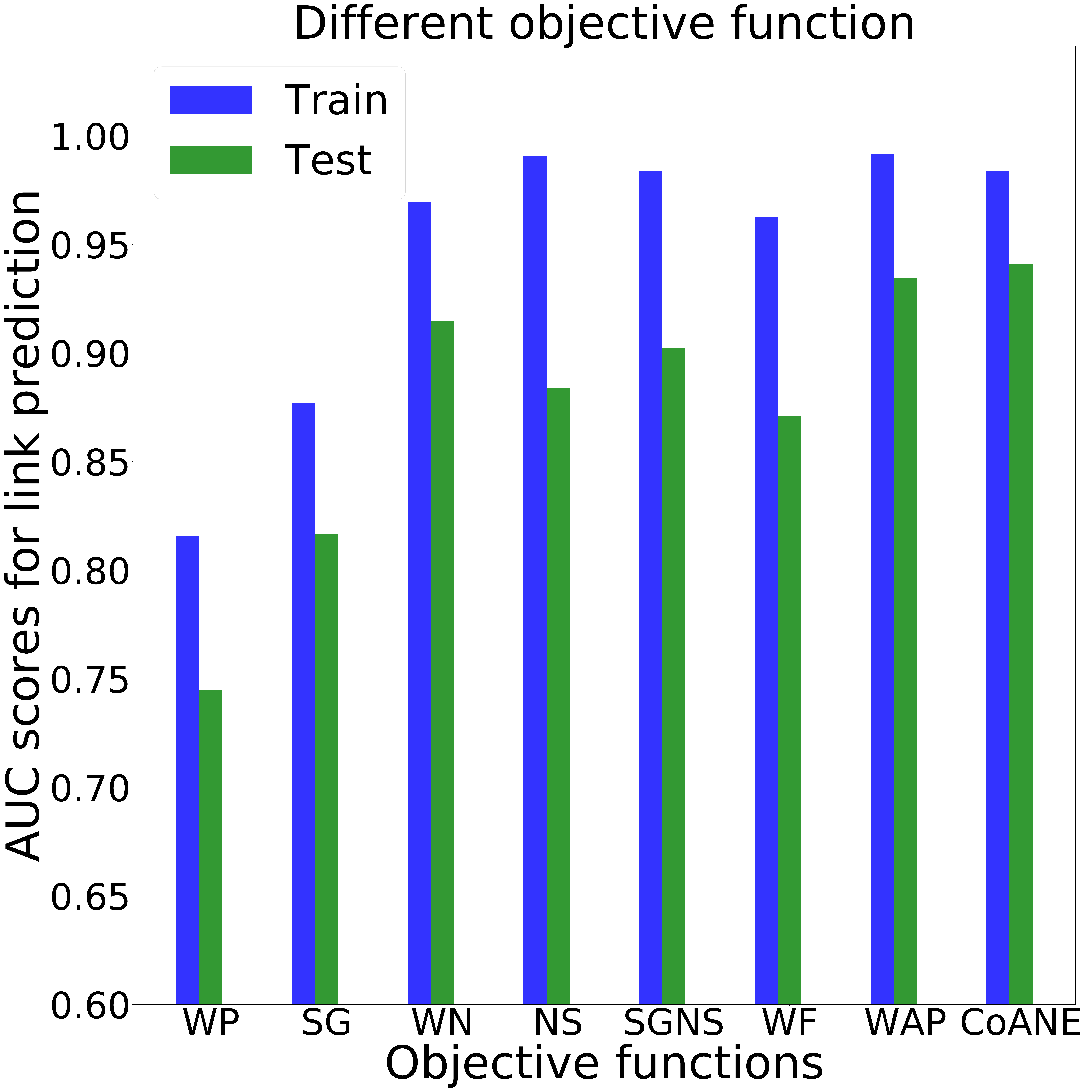}
         \caption{Effect of objective functions}
         \vspace{-2.5pt}
         \label{fig:P_obj}
     \end{subfigure}
      \hfill
     \begin{subfigure}[b]{0.24\textwidth}
         \centering
         \includegraphics[width=\textwidth]{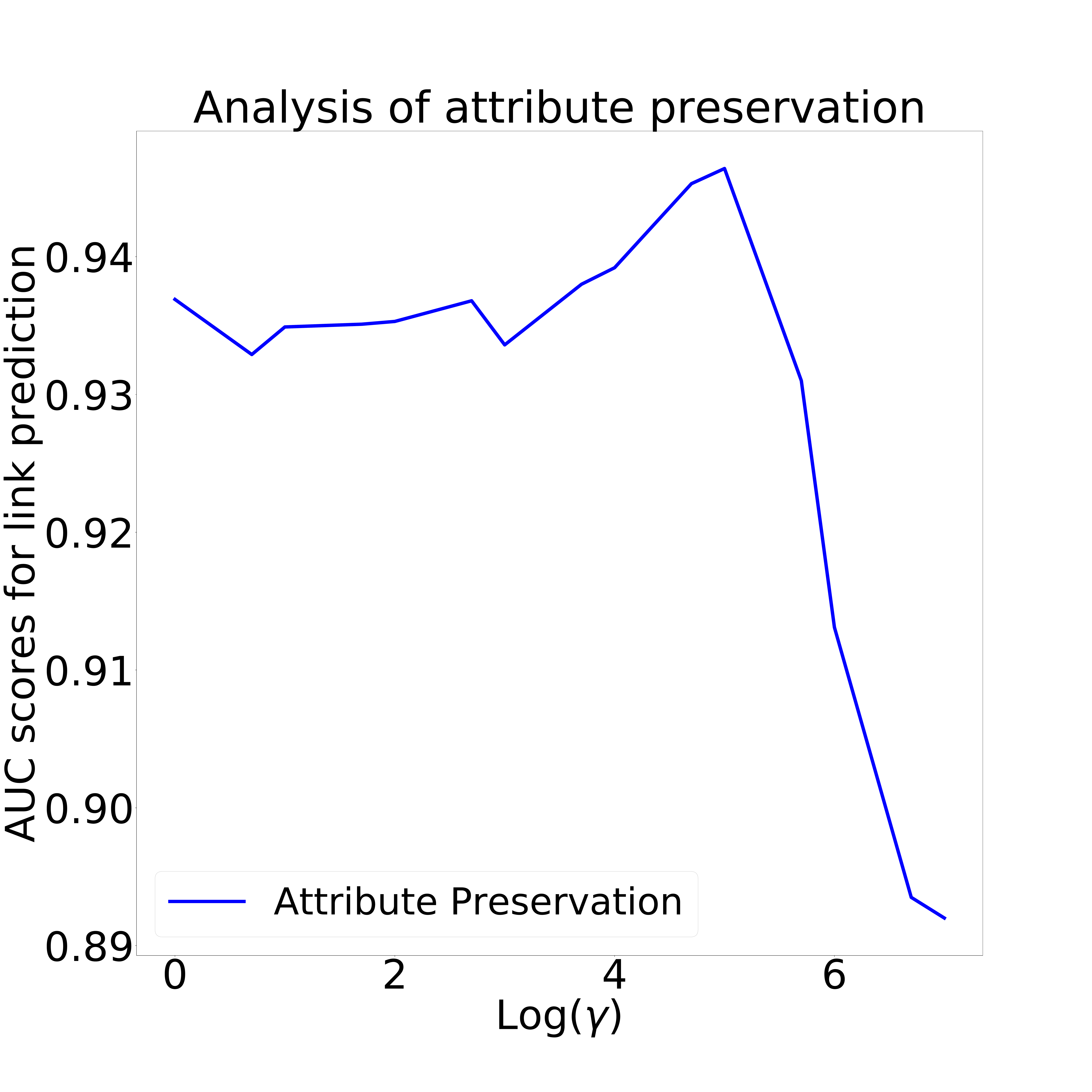}
         \caption{Attribute preservation.}
         \vspace{-2.5pt}
         \label{fig:P_Att}
     \end{subfigure}
        \caption{Analyzing each component in CoANE using Cora data.}
        \label{fig-pre}
        \vspace{-13pt}
\end{figure*}

\subsection{Discussion on CoANE Superiority}
\vspace{-1pt}
Experimental results show that CoANE leads to better and stable performance than existing solutions. The fundamental reason is the modeling of latent \textit{social circles} by the proposed multi-channel 1-D convolutional layer and the information preservation in CoANE. It can be further divided into two parts. First, CoANE can directly learn how subsets of neighbors sharing similar attributes can shape the potential \textit{social circles}, which cannot be well captured by joint structure-attribute learning-based models. Second, CoANE can give different distributions of weights to neighbors in the same hop and different hops away from the target node, which cannot be achieved by graph autoencoder-based models. Therefore, CoANE allows more flexibility and provides more deliberate learning in trainable parameters for fine-grained social circle modeling. The details are summarized as below.

Joint structure-attribute learning-based models, including DANE~\cite{Gao-Huang:IJCAI-2018}, ASNE~\cite{Liao-et-al:IEEE-2018}, and ANRL~\cite{zhang2018anrl}, first independently encode respective information (either graph structure or node attributes), then learn the correlation between structure and attributes, and have two prediction tasks: preserve the target/neighbor node(s) and reconstruct the attributes. However, in this way, the direct interactions between graph structure and node attributes cannot be captured. These methods simply treat node attributes as independent input, instead of viewing individual node attributes together with graph neighborhood. Therefore, they cannot learn the latent \textit{social circles} of the target node represented by how some neighbors possess similar attributes. 

Graph autoencoder models, including GAE/VGAE~\cite{Kipf-Welling:arXiv-2016} and ARGA/ARVGA~\cite{pan2018adversarially}, can simultaneously model network structure and node attributes. Besides, ANRL~\cite{zhang2018anrl} fuses the technologies of jointly feature learning and graph aggregation to preserve more information. However, their recursive scanning of same-hop neighbors from the first to higher order cannot capture the target node's \textit{social circles} (e.g., such as ``CS dept'', ``family'', and ``labmates''), which can span across different orders of neighbors. In other words, all neighbors in the same order are treated as having equal importance. They cannot distinguish same-hop neighbors from each other.

\subsection{Discussion of Information Preserved by CoANE}
We discuss the effectiveness with different designs of CoANE by considering its three main components, including (a) random walk, (b) convolutional mechanism, and (c) objective function. We apply link prediction on Cora dataset with the same settings as the previous evaluation, and exhibit the performance in the corresponding cases. 

(a) \textit{Random walk}. We pay attention to the contribution of our random walk and the original neighbor selection. First, we randomly choose a node and show the coverage of its neighbors in the network by depicting their paths on the t-SNE embedding plots. In Fig.~\ref{fig-pre1}, the asterisk represents the chosen node, and red lines are the real edges in the network. The paths in various colors in Fig.~\ref{fig:Pre_rw1} are contexts extracted by the random walk with the length of window $=5$, and the paths in blue in Fig.~\ref{fig:Pre_rw2} are the first two hop neighbors of the chosen node. We observe both of their regions cover most of the orange points but reach some nodes with other colors, especially for the random walk. This problem could be solved by paying different attention to each position in our CoANE model because these uncorrelated neighbors are located at the tail of the contexts. In addition, we can also find that the main region of the random walk paths (Fig. \ref{fig:Pre_rw1}) is more concentrating than the region of the first two hop neighbors (Fig. \ref{fig:Pre_rw2}). Such an effect implies that our random walk can help model better distinguish nodes that belong to the same clusters.
Second, we aim at presenting how these two neighbor selection cases affect the performance of link prediction. For fair and simplified comparison, we set the context length $= 1$ for the random walk case, and consider the first-hop neighbors as another case for comparison. In addition, we make the average number of generated contexts for two cases as close as possible by repeatedly generating the contexts. Eventually, we have $17.5$ and $22$ contexts per node for the cases of random walk and the first-hop neighbors, respectively. The results are displayed in Fig.~\ref{fig:P_rw1_CNN} (solid lines). The increment is obvious when using random walk contexts. Such a result indicates that the choice of nodes' neighbors is crucial for embedding learning.   

(b) \textit{Convolutional mechanism}. We discuss the preservation of positional information in the context by the learning layer comparison and the filter weights study. First, we compare the selection of feature extraction layer, including the convolution used by CoANE and the general mapping, i.e., fully connected (FC) layer. Applying the FC layer means that each node's features in the context are learned by the same parameters. The results are shown in Fig.~\ref{fig:P_rw1_CNN} (dashed lines). The convolutional layer leads to better performance and faster convergence, compared to the FC layer. The results imply that the positional information of nodes in the context should be considered. Since the neighborhood's network topology can to some extent be depicted by positional information in the generated contexts, the convolution exactly offers a more compatible mechanism to extract diverse positional properties. That is, CoANE can capture node features and their relationships by using only one convolutional layer.

Furthermore, we exhibit the weights of filters to explain the effectiveness of CNN. Although we have no knowledge about the attributes in the datasets, we expect that filters can give similar weights on specific attributes between the central node and its neighbors (contexts). The learned weights of CNN filters are shown in Fig.~\ref{fig:P_F}. In each subfigure, the x-axis represents the attribute dimensions, and the y-axis indicates the positions of the central node and its context nodes. Colors are used to display the weight values in the filters. To have a better observation, we sort the attributes dimensions of filters by the weights of central nodes. The top subfigure presents the weights of filters for all dimensions while the bottom subfigures show the weights in the bottom and the top $10$ dimensions. We can clearly find that the attribute weights of midst nodes with higher weights (violet) are often accompanied by higher weights of their neighbors. Such results imply that filters concentrate on similar attributes and the positional information in the context. These findings become more obvious when looking into the bottom and the top $10$ dimensions. Most neighbors of the midst have higher positive correlated weights; otherwise, and the rest have weights close to zero (cyan). In short, various filters are truly proficient in both searching features of their targets and giving weights according to the positions.

(c) \textit{Objective function}. We present the contribution of modeling context co-occurrence by our likelihood functions. Eight cases are discussed: (1) CoANE without positive graph likelihood (WP) (i.e., set $L_{pos}=0$), (2) using the general skip-gram model to replace positive graph likelihood (SG), i.e., simply computing dot product similarity for pairs of midst and neighbor, (3) CoANE without contextually negative sampling (WN) (i.e., set $L_{neg}=0$), (4) using general negative sampling to replace contextually negative sampling (NS), i.e., simply computing dot product similarity for a fixed number of negative samples via uniformly random selection, (5) combining skip-gram model with negative sampling to replace CoANE's positive and negative loss (SGNS) (i.e., (2) + (4)), (6) CoANE using the original network data without the attribute information (WF), (7) CoANE without the attribute preservation (WAP), and (8) the complete CoANE. We compare their performance in training and testing AUC using link prediction task. 

The results are shown in Fig.~\ref{fig:P_obj}. 
It can be clearly found that changing or removing any required components of CoANE brings performance damage. The effectiveness of each component of CoANE has been verified. By looking into the details, the worse AUC scores of WP and SG (i.e., without proper positive loss terms) prove the usefulness of our context co-occurrence matrices and positive graph likelihood.
The worse AUC scores of WN and NS (i.e., original settings of negative sampling) imply that it is effective to have our contextually negative sampling, and higher training scores and lower testing scores may indicate the potential overfitting of WN and NS. Contextually negative sampling can better deal with overfitting.
In addition, the AUC scores of combining skip-gram model and original negative sampling (SGNS) are not worse, comparing to the complete CoANE. We think it is because the model is still based on the proposed modeling of context co-occurrence that effectively distills features from network structure and node attributes.
Besides, the significant AUC difference between WF and the complete CoANE shows the contribution of attributes in embedding learning and the prediction task. 

For WAP, removing attribute preservation makes the model fit the training data well; nevertheless, some attributes can still benefit the model learning and improve the performance, as exhibited by the complete CoANE. We analyze the capability of attribute preservation by changing the attribute preservation controller $\gamma$ in Eq. (\ref{eq:AP}) based on the same experimental settings. We display the results in terms of AUC scores by varying $\log(\gamma)$ in Fig.~\ref{fig:P_Att}. It can be found that the curve is first increasing, and then goes down as the increment of $\log(\gamma)$. The reason is that much smaller attribute preservation does not affect the model learning; however, larger $\gamma$ values would greatly dominate the embedding learning, i.e., focusing more on attribute preservation and weakening structure learning. In the setting with $\log(\gamma) = 5$ (i.e., $1e5$), the attribute preservation can make CoANE achieve better performance.
\vspace{-6pt}

\section{Conclusion}
\vspace{-2pt}
This paper proposes a novel context co-occurrence-aware attributed network embedding, CoANE. 
The main idea is to preserve three-fold information, the network structure, node attributes, and distilling the proper convolution between network structure and node attributes through specific contexts. 
When applying the embeddings generated by CoANE, we prove that the performance conducted on three essential network analysis tasks, including link prediction, node label classification, and node clustering, can get significantly and consistently boosted across five real datasets, comparing to state-of-the-art competing methods.  Such results clearly exhibit the effectiveness of CoANE. A number of advanced analyses on filters' weights, parameter sensitivity, and model contributions provide a robust experimental study, and the results unfold where the superiority of CoANE comes from.

\vspace{-4pt}


\ifCLASSOPTIONcompsoc
  \section*{Acknowledgments}
\else
  \section*{Acknowledgment}
\fi

This work is supported by Ministry of Science and Technology (MOST) of Taiwan under grants 109-2636-E-006-017 (MOST Young Scholar Fellowship) and 109-2221-E-006-173, and also by Academia Sinica under grant AS-TP-107-M05.

\vspace{-4pt}

\bibliographystyle{plain}
\bibliography{CoANE}

\vspace{-40pt}
\begin{IEEEbiography}[{\includegraphics[width=1in,height=1.25in,clip,keepaspectratio]{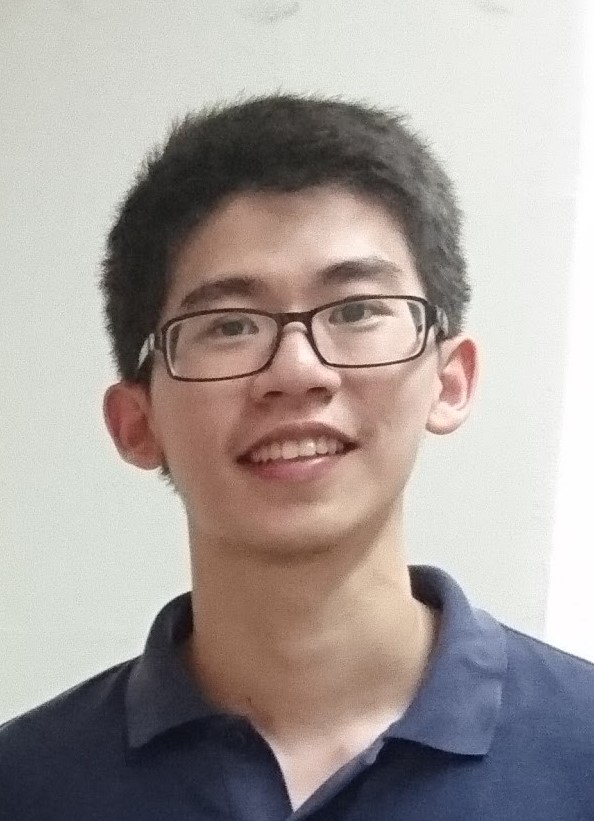}}]%
{I-Chung Hsieh} received a Master's degree in Science in Statistical Science from National Chung Cheng University, Chiayi, Taiwan, in 2017. He is a research assistant in Networked Artificial Intelligence Laboratory at National Cheng Kung University, Tainan, Taiwan from 2018. His main research interests include Privacy protection on Graph, Graph representation learning, Data Mining, and Deep Learning. Now, his current research has been accepted and presented at NeurIPS GRL 2019.
\end{IEEEbiography}
\vspace{-35pt}
\begin{IEEEbiography}[{\includegraphics[width=1in,height=1.25in,clip,keepaspectratio]{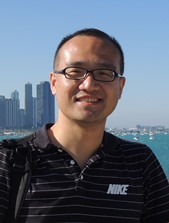}}]%
{Cheng-Te Li} is an Associate Professor at Institute of Data Science and Department of Statistics, National Cheng Kung University (NCKU), Tainan, Taiwan. He received my Ph.D. degree (2013) from National Taiwan University. 
Before NCKU, He was an Assistant Research Fellow (2014-2016) at CITI, Academia Sinica. 
Dr. Li's research targets at Machine Learning, Deep Learning, Data Mining, Social Networks and Social Media, Recommender Systems, and Natural Language Processing. He has a number of papers published at top conferences, including KDD, WWW, ICDM, CIKM, SIGIR, IJCAI, ACL, EMNLP, NAACL, RecSys, and ACM-MM. 
\end{IEEEbiography}
\end{document}